\documentclass[11pt]{article}
\usepackage{amsfonts}

\usepackage{graphics}
\usepackage{indentfirst}
\usepackage{cite}
\usepackage{latexsym}
\usepackage{amsmath}
\usepackage{amssymb}
\usepackage[dvips]{epsfig}
\usepackage{amscd}

\newtheorem{theorem}{Theorem}[section]
\newtheorem{remark}{Remark}[section]

\newtheorem{lemma}[theorem]{Lemma}
\newtheorem{pro}{Proposition}[section]

\renewcommand{\div}{ {\rm div }  }

\def\tp{ {P(\tn)}}

\newcommand{\pa}{\partial}
\renewcommand{\r}{\mathbb{R}}

\newcommand{\ia}{\int_0^T}

\newcommand{\bt}{\begin{theorem}}
\newcommand{\bl}{\begin{lemma}}
\newcommand{\el}{\end{lemma}}
\newcommand{\et}{\end{theorem}}
\newcommand{\ga}{\gamma}

\newcommand{\al}{\alpha}
\newcommand{\de}{\delta}
\newcommand{\ve}{\varepsilon}
\newcommand{\la}{\label}

\newcommand{\ol}{\overline}

\newcommand{\bn}{\begin{eqnarray}}
\newcommand{\en}{\end{eqnarray}}
\newcommand{\bnn}{\begin{eqnarray*}}
\newcommand{\enn}{\end{eqnarray*}}

\newcommand{\bnnn}{\begin{eqnarray*}}
\newcommand{\ennn}{\end{eqnarray*}}

\newcommand{\ba}{\begin{aligned}}
\newcommand{\ea}{\end{aligned}}
\newcommand{\be}{\begin{equation}}
\newcommand{\ee}{\end{equation}}
\def\O{{\r^3}}
\def\p{\partial}
\def\norm[#1]#2{\|#2\|_{#1}}

\def\o{\omega}

\newcommand{\no}{\nonumber\\}

\newcommand{\n}{\rho}
\newcommand{\si}{\sigma}

\def\la{\label}
\def\pp{P-P(\tn)}
\def\na{\nabla}
\def\on{\bar\rho}

\def\tn{\tilde{\rho}}

\makeatletter      
\@addtoreset{equation}{section}
\makeatother       

\title{Global Well-Posedness of Classical Solutions with Large
Oscillations and Vacuum to the Three-Dimensional Isentropic
Compressible Navier-Stokes Equations\thanks{This research is
supported in part by Zheng Ge Ru Foundation, Hong Kong RGC
Earmarked Research Grants CUHK4040/06P and CUHK4042/08P, and a
Focus Area Grant from The Chinese University of Hong Kong. The
research of J. Li is partially supported by
 NSFC Grant No. 10971215.
 Email: xdhuang@ustc.edu.cn (X. Huang), ajingli@gmail.com (J. Li),
zpxin@ims.cuhk.edu.hk (Z. Xin).}}

\author{Xiangdi H{\small UANG}$^{a,c}$,  Jing L{\small I}$^{b,c}$, Zhouping X{\small IN}$^c$
\\[3mm] {\normalsize $^a$ Department of Mathematics,} \\
{\normalsize University of Science and Technology of China, Hefei
230026, P. R. China} \\[2mm]
{\normalsize $^b$ Institute of Applied Mathematics, AMSS,} \\
{\normalsize Academia Sinica, Beijing 100190,
P. R. China}\\[2mm]
{\normalsize $^c$ The Institute of Mathematical Sciences,} \\
{\normalsize The Chinese University of Hong Kong, Shatin, Hong
Kong}}

\date{}

\begin{document}
\maketitle
\begin{abstract}
We establish the global existence and uniqueness of classical
solutions to the Cauchy problem for the isentropic compressible
Navier-Stokes equations in three spatial dimensions with smooth
initial data which are of small energy but possibly large
oscillations with constant state as far field which could be
either vacuum or non-vacuum. The initial density is allowed to
vanish and the spatial measure of the set of vacuum can be
arbitrarily large, in particular, the initial density can even
have compact support. These results generalize previous results on
classical solutions for initial densities being strictly away from
vacuum, and are the first for global classical solutions which may
have large oscillations and can contain vacuum states.
\end{abstract}

\section{Introduction}
The time evolution of the density and the velocity of a general
viscous   isentropic compressible   fluid occupying a domain
$\Omega\subset \r^3$ is governed by the   compressible Navier-Stokes
equations:
   \bn \la{a1}  \begin{cases}\n_t+{\rm div} (\n u)=0,\\
 (\n u)_t+{\rm div}(\n u\otimes u)-\mu\Delta u-(\mu+\lambda)\na({\rm
div} u)+\na P(\n)=0, \end{cases}\en where $\rho\ge 0,$
$u=(u^1,u^2,u^3)$   and $P=a\n^\ga( a>0, \ga>1)  $ are the fluid
density, velocity and pressure, respectively.   The constant
viscosity coefficients $\mu$ and $\lambda$  satisfy the physical
restrictions: \be\la{h3} \mu>0,\quad \mu + \frac{3}{2}\lambda\ge 0.
\ee
 Let $\Omega=\r^3 $ and    $\tn$ be a fixed nonnegative constant.
 We look for the solutions, $(\n(x,t),u(x,t)),$ to the Cauchy problem
for (\ref{a1}) with  the far field behavior:\bn\la{h1}
u(x,t)\rightarrow 0,\quad \rho(x,t)\rightarrow\tn\ge 0,  \quad\mbox{
as }\, |x|\rightarrow\infty, \en and initial data, \be \la{h2}
(\rho,u)|_{t=0}=(\rho_0,u_0),\quad x\in \r^3.  \ee

There are huge literatures on the large time existence and
behavior of solutions to (\ref{a1}). The one-dimensional problem
has been studied extensively by many people, see
\cite{Kaz,Ser1,Ser2,Hof} and the references therein. For the
multi-dimensional case, the local existence and uniqueness of
classical solutions are known in \cite{Na,se1}  in the absence of
vacuum and recently, for strong solutions also, in \cite{K3,K1,
K2, S2} for the case that the initial density need not be positive
and may vanish in open sets. The global classical solutions were
first obtained by Matsumura-Nishida \cite{M1} for initial data
close to a non-vacuum equilibrium in some Sobolev space $H^s.$ In
particular, the theory requires that the solution has small
oscillations from a uniform non-vacuum state so that the density
is strictly away from the vacuum and the gradient of the density
remains bounded uniformly in time. Later, Hoff \cite{H3,Hof2}
studied the problem for discontinuous initial data. For the
existence of solutions for arbitrary data (the far field density
is vacuum, that is, $\tn=0$), the major breakthrough is due to
Lions \cite{L1} (see also  Feireisl \cite{F1}), where he obtains
global existence of weak solutions - defined as solutions with
finite energy - when the exponent $\ga$ is suitably large. The
main restriction on initial data is that the initial energy is
finite, so that the density vanishes at far fields, or even has
compact support. However, little is known on the structure of such
weak solutions. Recently, under the additional assumptions that
the viscosity coefficients $\mu $ and $\lambda$ satisfy \bn
\la{ua6} \mu>\max\{ 4\lambda,-\lambda\},\en and for the
 far field density away from vacuum $(\tn>0),$  Hoff (\cite{Ho3,hs,ht})
obtained a new type of global weak solutions with small energy,
which have extra regularity information compared with those large
weak ones constructed by Lions (\cite{L1}) and Feireisl (\cite{F1}).
Note that here the weak solutions may contain vacuum though the
spatial measure of the set of vacuum has to be small. Moreover,
under some additional conditions which prevent  the appearance of
vacuum states in the data, Hoff (\cite{Ho3, ht}) obtained also
classical solutions.

It should be noted that in the presence of vacuum, the global
well-posedness of classical solutions and the regularity and
uniqueness of those weak solutions (\cite{L1,F1,Ho3}) remains
completely open. Indeed, this is a subtle issue since, in general,
one would not expect such general results due to Xin's blow-up
results   in \cite{X1}, where it is shown that in the case that the
initial density has compact support, any smooth solution to the
Cauchy problem of the non-barotropic compressible Navier-Stokes
system without heat conduction blows up in finite time for any space
dimension, and the same holds for the isentropic case (\ref{a1}), at
least in one-dimension, and the symmetric two-dimensional case
(\cite{hlx3}). See also the recent generalizations to the cases for
the non-barotropic compressible Navier-Stokes system with  heat
conduction (\cite{cj}) and for non-compact but rapidly decreasing at
far field initial densities (\cite{R}).

   In this paper, we will study the global existence and uniqueness
   of classical   solutions  to the Cauchy problem for the isentropic
   compressible
Navier-Stokes equations, (\ref{a1}), in three-dimensional space with
smooth initial data which are of small energy but possibly large
oscillations with constant state as   far field
 which could be either vacuum $(\tn=0)$ or non-vacuum  $(\tn>0);$
 in particular, the
initial density is allowed to vanish,
even has compact support.

Before stating the main results, we explain the notations and
conventions used throughout this paper. We denote
$$\int fdx=\int_{\r^3}fdx.$$ For $1\le r\le \infty $ and $\beta>0,$ we denote the standard
homogeneous and inhomogeneous Sobolev spaces as follows:
   \bnnn \begin{cases}L^r=L^r(\r^3),\quad D^{k,r}  =
   \left.\left\{u\in
L^1_{loc}(\r^3)\,\right|\|{\nabla^k u}\|_{L^r}<\infty\right\},
\quad \norm[D^{k,r} ]{u}\triangleq  \|{\nabla^k u}\|_{L^r},\\
W^{k,r}  = L^r \cap D^{k,r} , \quad H^k = W^{k,2} ,\quad D^k  =
D^{k,2} ,
 \quad  D^1  = \left. \left\{u\in L^6 \,\right|
 \|{\nabla u}\|_{L^2}<\infty \right\}\\ \dot H^\beta=\left\{f:\r^3
 \rightarrow \r\left|\|f\|^2_{\dot H^\beta}=
 \displaystyle{\int} |\xi|^{2\beta}|\hat f(\xi)|^2d\xi<\infty\right.
 \right\} ,\end{cases}\ennn where $\hat f$ is the Fourier transform
 of $f.$

The initial energy is defined as: \be\la{h5} C_0 =
\int\left(\frac{1}{2}\rho_0|u_0|^2 + G(\rho_0)\right)dx, \ee where
$G$ denotes the potential energy density given by \bnn
G(\n)\triangleq \n\int_{\tn}^\n\frac{P(s)-P(\tn)}{s^2}ds.\enn It is
clear that \bnn
\begin{cases}
G(\n)=\frac{1}{\ga-1}P,& \mbox{ if }\quad\tn=0,\\
c_1(\on,\tn)(\n-\tn)^2\le G(\n)\le c_2(\on,\tn)(\n-\tn)^2,&\mbox{ if
}\quad\tn>0,\,\,0\le\n\le\on,\end{cases}\enn for   positive
constants $c_1(\on,\tn)$ and $c_2(\on,\tn).$

Then the main results in this paper  can be stated  as follows:
\begin{theorem}\la{th1} Assume that (\ref{h3}) holds.
  For given numbers $M>0$ (not necessarily small), $\beta\in (1/2,1],$
  and $\on\ge\tn+1,$ suppose that
    the initial data $(\n_0,u_0)$ satisfy
\be \la{co1} \n_0|u_0|^2+G(\n_0)\in L^1,\quad  u_0\in \dot
H^\beta\cap D^1 \cap D^3  ,\quad (\rho_0-\tn,P(\n_0)-P(\tn))\in H^3
,  \ee
 \be\la{h7} 0\le\inf\rho_0\le\sup\rho_0\le\bar{\rho},\quad
   \|u_0\|_{\dot H^\beta} \le M,   \ee
and the compatibility condition \be\la{co2} -\mu\triangle u_0
-(\mu+\lambda)\nabla\text{div}u_0 + \nabla P(\rho_0) = \rho_0g, \ee
for some $g\in D^1 $ with $\rho_0^{1/2}g\in L^2.$
 Then there exists a positive constant $\ve$ depending
 on $\mu,\lambda,\tn, a, \ga, $  $\on, \beta$ and $M$  such that if
 \be\la{i7}
     C_0\le\ve,
   \ee  the Cauchy problem
  (\ref{a1}) (\ref{h1}) (\ref{h2})
  has a unique global classical solution $(\rho,u)$ in
   $\O\times(0,\infty)$ satisfying for
  any $0<\tau<T<\infty,$
  \be\la{h8}
  0\le\rho(x,t)\le 2\bar{\rho},\quad x\in \r^3,\, t\ge 0,
  \ee
   \be
   \la{h9}\begin{cases}
  (\rho-\tn,P-P(\tn))\in C([0,T];H^3), \\
  u\in C([0,T];D^1\cap D^3)\cap L^2(0,T;D^4)
\cap L^\infty(\tau,T;D^4),\\
  u_t\in L^{\infty}(0,T;D^1)\cap L^2(0,T;D^2)\cap
L^{\infty}(\tau,T;D^2)\cap H^1(\tau,T;D^1),\\   \sqrt{\n}u_t\in
L^\infty(0,T;L^2) ,\end{cases} \ee
  and the following large-time behavior:
  \be\la{h11}
  \lim_{t\rightarrow\infty}\int(|\rho-\tn|^q + \rho^{1/2}|u|^4+
  |\nabla u|^2)(x,t)dx = 0,  \ee
    for all  \bn\la{eq1} q\in\begin{cases}  (2 ,\infty),
    \quad \mbox{ for }\tn>0,\\
      (\ga ,\infty)  ,\quad\mbox{ for }\tn=0.
    \end{cases}\en

\end{theorem}

Similar to our previous studies on the Stokes approximation
equations in \cite{lx}, we can obtain from (\ref{h11}) the
following large time behavior of the gradient of the density when
vacuum states appear initially and the far field density is away
from vacuum, which is completely in contrast to the classical
theory (\cite{ht,M1}).
\begin{theorem} \la{th2}
In addition to the conditions  of Theorem \ref{th1}, assume further
that there exists some point $x_0\in \r^3$ such that $\rho
_0(x_0)=0.$ Then if $\tn>0,$ the unique global    classical solution
$(\rho,u)$ to the Cauchy problem (\ref{a1}) (\ref{h1}) (\ref{h2})
obtained in Theorem \ref{th1} has to blow up as $t\rightarrow
\infty,$ in the sense that for any $r>3,$
$$\lim\limits_{t\rightarrow \infty}\|\nabla \rho(\cdot,t)
\|_{L^r }=\infty.$$
\end{theorem}

A few remarks are in order:

\begin{remark}  The solution obtained in Theorem \ref{th1} becomes
a classical one for positive time. Although it has small energy, yet
whose oscillations could be arbitrarily large. In particular, both
interior and far field vacuum states are allowed.
\end{remark}

\begin{remark} In the case that the far field density is away from vacuum,
i.e., $\tn>0,$ the conclusions in Theorem \ref{th1} generalize the
classical theory of Matsumura-Nishida (\cite{M1}) to the case of
large oscillations since in this case, the requirement of small
energy, (\ref{i7}), is equivalent to smallness of the mean-square
norm of $(\n_0-\tn,u_0).$ However, though the large-time
asymptotic behavior (\ref{h11}) is similar to that in \cite{M1},
yet our solution may contain vacuum states, whose appearance leads
to the large time blowup behavior stated in Theorem \ref{th2},
this is in sharp contrast to that in \cite{M1,ht} where the
gradients of the density are suitably small uniformly for all
time.
\end{remark}

\begin{remark}
When the far field density is   vacuum, i.e., $\tn=0,$  the small
energy assumption, (\ref{i7}), is equivalent to that both the
kinetic energy and the total pressure are suitably small. There is
no requirement on the size of the set of vacuum states. In
particular, the initial density may have compact support. Thus,
Theorem \ref{th1} can be regarded as a uniqueness and
 regularity
theory of Lions-Feireisl's weak solutions  in \cite{L1,F1} with
small initial energy. It should also be noted that the conclusions
in Theorem \ref{th1} for the case of $\tn=0$ are somewhat surprising
since for the isentropic compressible Navier-Stokes equations
(\ref{a1}), any non-trivial one-dimensional smooth solution with
initial compact supported density blows up in finite time
(\cite{X1}), and the same holds true for two-dimensional smooth
spherically symmetric solutions (\cite{hlx3}).

\end{remark}

\begin{remark} It should be emphasized that in Theorem \ref{th1},
the viscosity coefficients are only assumed to satisfy the physical
conditions  (\ref{h3}). While the theory on weak small energy
solutions, developed in
 \cite{Ho3,ht}, requires the additional assumption (\ref{ua6})
 which is crucial in establishing the time-independent upper bound for the
 density in the arguments  in  \cite{Ho3,ht}.
\end{remark}

\begin{remark} For the incompressible Navier-Stokes system, a lot of
results on the global wellposedness in scaling invariant  spaces are
available \cite{fk,kato,koch}.  In particular, Fujita-Kato \cite{fk}
and Kato \cite{kato} proved that the system   is globally wellposed
for small initial data in the homogeneous Sobolev spaces $\dot
H^{1/2}$ or in $L^3.$ In our case,    the initial energy is small,
therefore, we need the boundedness assumptions on the $\dot
H^{\beta}$-norm of the initial velocity. It should be noted here
that $\dot H^{\beta}\hookrightarrow L^{6/(3-2\beta)}$ and
$6/(3-2\beta)>3$ for $\beta>1/2,$ which implies that, compared with
the   results in \cite{fk,kato}, our conditions on the initial
velocity may be optimal under the smallness conditions on the
initial energy.
\end{remark}

\begin{remark}   Similar ideas can be applied to study the case
on bounded domain. This will be reported in a forthcoming paper
\cite{hlx4}.
\end{remark}

We now comment on the analysis of this paper. Note that for initial
data in the class satisfying (\ref{co1})-(\ref{co2}) except $u_0\in
\dot H^\beta,$ the local existence and uniqueness of  classical
solutions to the Cauchy problem, (\ref{a1})-(\ref{h2}), have been
established recently in \cite{K3}. Thus, to extend the classical
solution globally in time, one needs global a priori estimates on
smooth solutions to (\ref{a1})-(\ref{h2}) in suitable higher norms.
Some of the main new difficulties are due to the appearance of
vacuum and that there are no other constraints on the viscosity
coefficients beyond the physical conditions (\ref{h3}). It turns out
that the key issue in this paper is to derive both the
time-independent upper bound for the density and the time-depending
higher norm estimates of the smooth solution $(\rho, u)$. We start
with the basic energy estimate and the initial layer analysis, and
succeed in deriving an estimate on the spatial weighted $L^3$-norm
of the velocity,   the weighted spatial mean estimates on  both the
gradient   and the material derivatives of the velocity. This is
achieved by modifying the basic elegant estimates on the material
derivatives of the velocity developed by Hoff
(\cite{H3,Ho3,hof2002}) in the theory of small energy weak solutions
with non-vacuum far fields and an interpolation argument. Then we
are able to obtain the desired estimates on
$L^1(0,\min\{1,T\};\,L^\infty({\r}^3))$-norm
 and the time-independent ones on
   $L^{8/3}(\min\{1,T\},T;\,L^\infty({\r}^3))$-norm
   of the effective viscous flux (see (\ref{hj1}) for the definition).
It follows from these key estimates and Zlotnik's inequality (see
Lemma \ref{le1}) that the density admits a time-uniform upper
bound which is the key for global estimates of classical
solutions. This approach to estimate a uniform upper bound for the
density is motivated by our previous analysis on the
two-dimensional Stokes approximation equations in \cite{lx}. The
next main step is to bound the gradients of the density and the
velocity. Motivated by our recent studies (\cite{hlx1,hlx,hx2}) on
the blow-up criteria of classical (or strong) solutions to
(\ref{a1}), such bounds can be obtained by solving a logarithm
Gronwall inequality based on a Beal-Kato-Majda type inequality
(see Lemma \ref{le9}) and the a priori estimates we have just
derived, and  moreover, such a derivation yields simultaneously
also the bound for $L^1(0,T;L^\infty({\r}^3))$-norm of the
gradient of the velocity, see Lemma \ref{le11} and its proof. It
should be noted here that we do not require smallness of the
gradient of the initial density which prevents the appearance of
vacuum (\cite{ht,M1}). Finally, with these a priori estimates on
the gradients of the density and the velocity  at hand, one can
estimate the higher order derivatives by using the same arguments
as in \cite{hx2} to obtain the desired results.

The rest of the paper is organized as follows: In Section 2, we
collect some elementary facts and inequalities which will be needed
in later   analysis. Section 3 is devoted to deriving the necessary
a priori estimates on classical solutions which are needed to extend
the local solution to all time. Then finally, the main results,
Theorem 1.1 and Theorem 1.2, are proved in Section 4.

\section{Preliminaries}\la{se2}

In this section, we will recall some  known facts and elementary
inequalities which will be used frequently later.

We start with the local existence and uniqueness of classical
solutions when the initial density may not be positive and may
vanish in an open set.
\begin{lemma}[\cite{K3}]\la{th0} For $\tn\ge 0,$ assume that
 the initial data $(\n_0\ge 0 ,u_0)$ satisfy  (\ref{co1})-(\ref{co2})
  except $u_0\in \dot H^{\beta}.$
   Then
there exist  a small time $T_*>0$ and a unique classical solution
$(\rho, u)$ to the Cauchy problem  (\ref{a1}) (\ref{h1}) (\ref{h2})
on $\O\times(0,T_*]$ such that \be \la{ba1}\begin{cases}
  (\rho-\tn,P-P(\tn))\in C([0,T_*];H^3), \\  u\in C([0,T_*];D^1\cap D^3)\cap L^2(0,T_*;D^4),\\
  u_t\in L^{\infty}(0,T_*;D^1)\cap L^2(0,T_*;D^2),\quad
\sqrt{\rho}u_{ t}\in L^\infty(0,T_*;L^2), \\  \sqrt{\rho}u_{tt}\in
L^2(0,T_*;L^2), \quad t^{1/2}u \in L^\infty(0,T_*;D^4), \\
t^{1/2}\sqrt{\n}u_{tt} \in L^\infty(0,T_*;L^2),\quad t u_t \in
L^\infty(0,T_*;D^3),\\   t u_{tt} \in L^\infty(0,T_*;D^1)\cap
L^2(0,T_*;D^2) .\end{cases}\ee
\end{lemma}

Next, the following well-known Gagliardo-Nirenberg inequality
  will be used later frequently (see \cite{la}).

\begin{lemma}
[Gagliardo-Nirenberg]\la{l1} For  $p\in [2,6],q\in(1,\infty), $ and
$ r\in  (3,\infty),$ there exists some generic
 constant
$C>0$ which may depend  on $q,r$ such that for   $f\in H^1({\r^3}) $
and $g\in L^q({\r^3})\cap D^{1,r}({\r^3}), $    we have\bn
\la{g1}\|f\|_{L^p}^p\le C \|f\|_{L^2}^{(6-p)/2}\|\na
f\|_{L^2}^{(3p-6)/2} ,\en  \bn
\la{g2}\|g\|_{C\left(\ol{\r^3}\right)} \le C
\|g\|_{L^q}^{q(r-3)/(3r+q(r-3))}\|\na g\|_{L^r}^{3r/(3r+q(r-3))} .
\en
\end{lemma}

We now state some elementary estimates which follow from (\ref{g1})
and the standard $L^p$-estimate  for the following elliptic system
derived from the momentum equations in (\ref{a1}): \be\la{h13}
\triangle F = \text{div}(\rho\dot{u}),\quad\mu \triangle \o =
\nabla\times(\rho\dot{u}), \ee where \be \la{hj1}\dot f\triangleq
f_t+u\cdot\nabla f,\quad F\triangleq(2\mu + \lambda)\text{div}u -
P(\rho) + P(\tn),\quad\o \triangleq\na\times u, \ee are  the
material derivative of $f,$ the effective viscous flux and the
vorticity respectively.

\begin{lemma} \la{le3}
  Let $(\rho,u)$ be a smooth solution of
   (\ref{a1}) (\ref{h1}).
    Then there exists a generic positive
   constant $C$ depending only on $\mu$ and $\lambda$ such that for any $p\in [2,6]$
          \bn\la{h19}
   &&\|{\nabla F}\|_{L^p} + \|{\nabla \o}\|_{L^p}
   \le C\norm[L^p]{\rho\dot{u}},\\
   &&  \la{h20}\norm[L^p]{F} + \norm[L^p]{\o}
   \le C \norm[L^2]{\rho\dot{u}}^{(3p-6)/(2p) }
   \left(\norm[L^2]{\nabla u}
   + \norm[L^2]{P-P(\tn)}\right)^{(6-p)/(2p)} ,\quad\quad\quad\quad
\\&&\la{h18}
   \norm[L^p]{\nabla u} \le C \left(\norm[L^p]{F} + \norm[L^p]{\o}\right)+
   C \norm[L^p]{P-P(\tn)},
  \\&&\la{h17} \|\na u\|_{L^p}\le C   \norm[L^2]{\nabla
u}^{(6-p)/(2p)}\left(\norm[L^2]{\rho\dot{u}}+\norm[L^6]{P-P(\tn)}
\right)^{(3p-6)/(2p) }.
  \en
\end{lemma}
{\it Proof.} The standard $L^p$-estimate for the elliptic system
(\ref{h13}) yields directly (\ref{h19}), which, together with
(\ref{g1}) and (\ref{hj1}), gives (\ref{h20}).

Note that $-\Delta u=-\na {\rm div}u +\na\times \o,$ which implies
that \bnn\na u=-\na(-\Delta)^{-1}\na {\rm
div}u+\na(-\Delta)^{-1}\na\times \o.\enn Thus the standard $L^p$
estimate shows that \bnn  \|\na u\|_{L^p}\le C (\|{\rm
div}u\|_{L^p}+\|\o\|_{L^p}),\,\, \mbox{ for }p\in [2,6],\enn which,
together with (\ref{hj1}), gives (\ref{h18}). Now (\ref{h17})
follows from (\ref{g1}), (\ref{h18}) and (\ref{h19}).

Next, the following Zlotnik  inequality will be used to get the
uniform (in time) upper bound of the density $\n.$
\begin{lemma}[\cite{zl1}]\la{le1}   Let the function $y$ satisfy
\bnn y'(t)= g(y)+b'(t) \mbox{  on  } [0,T] ,\quad y(0)=y^0, \enn
with $ g\in C(R)$ and $y,b\in W^{1,1}(0,T).$ If $g(\infty)=-\infty$
and \be\la{a100} b(t_2) -b(t_1) \le N_0 +N_1(t_2-t_1)\ee for all
$0\le t_1<t_2\le T$
  with some $N_0\ge 0$ and $N_1\ge 0,$ then
\bnn y(t)\le \max\left\{y^0,\overline{\zeta} \right\}+N_0<\infty
\mbox{ on
 } [0,T],
\enn
  where $\overline{\zeta} $ is a constant such
that \be\la{a101} g(\zeta)\le -N_1 \quad\mbox{ for }\quad \zeta\ge \overline{\zeta}.\ee
\end{lemma}

Finally, we state the following Beal-Kato-Majda type inequality
which was proved in \cite{bkm} when $\div u\equiv 0$ and will be
used later to estimate $\|\nabla u\|_{L^\infty}$ and
$\|\nabla\rho\|_{L^2\cap L^6}$.
\begin{lemma}  \la{le9}  For $3<q<\infty,$ there is a
constant  $C(q)$ such that  the following estimate holds for all
$\na u\in L^2(\O)\cap D^{1,q}({\r^3}),$ \be\la{ww7}\ba \|\na
u\|_{L^\infty({\r^3})}&\le C\left(\|{\rm div}u\|_{L^\infty({\r^3})}+
\|\o\|_{L^\infty({\r^3})} \right)\log(e+\|\na^2
u\|_{L^q({\r^3})})\\&\quad+C\|\na u\|_{L^2(\O)} +C . \ea\ee
\end{lemma}

{\it Proof.} The proof is similar to that of (15) in \cite{bkm} and
is sketched here for completeness. It follows from the Poisson's
formula that \be\ba u (x)&=-\frac{1}{4\pi}\int \frac{\Delta
u(y)}{|x-y|}dy\no&\equiv \int  {\rm div}u(y)K(x-y)dy-\int
K(x-y)\times \o (y)dy\\&\triangleq v+w,\ea\ee where
$$K(x-y)\triangleq\frac{ x-y }{4\pi|x-y|^3},$$ satisfies
\be \la{ww1}|  K(x-y)\le C|x-y|^{-2}, \quad |\na K(x-y)|\le
C|x-y|^{-3}.\ee

It suffices to estimate the term $\na v$ since $\na w$ can be
handled similarly (see \cite{bkm}). Let $\de\in(0,1]$ be a constant
to be chosen and introduce a cut-off function $\eta_\de(x)$
satisfying $\eta_\de(x)=1$ for $|x|<\de,\eta_\de(x)=0$ for
$|x|>2\de,$ and $|\na\eta_\de(x)|\le C\de^{-1}.$ Then $\nabla v$ can
be rewritten as \be\la{ww2}\ba \na v&= \int \eta_\de(y)K(y)\na{\div
u}(x-y)dy-\int \na\eta_\de(x-y) K(x-y){\div u}(y)  dy\\& \quad+ \int
(1-\eta_\de(x-y))\na K(x-y){\div u}(y) dy  .\ea\ee Each term on the
righthand side of (\ref{ww2}) can be estimated by (\ref{ww1}) as
follows: \be\la{ww3}
\ba  & \left|\int \eta_\de(y)K(y)\na{\div u}(x-y)dy\right|\\
&\le C\| \eta_\de(y)K(y)\|_{L^{q/(q-1)}}\|\na^2u\|_{L^q}\\ &\le C
\left(\int_0^{2\de}r^{-2q/(q-1)}r^2dr\right)^{(q-1)/q}\|\na^2u\|_{L^q}\\
&\le C\de^{(q-3)/q}\|\na^2u\|_{L^q},\ea\ee \be \la{ww4}\ba
 & \left|\int \na\eta_\de(x-y) K(x-y){\div u}(y) dy\right|\\&\le
\int|\na\eta_\de(z)| |K(z)|dz \|{\div u}\|_{L^\infty} \\&\le
C\int_\de^{2\de}\de^{-1}r^{-2}r^2dr \|{\div u}\|_{L^\infty} \\&\le C
\|{\div u}\|_{L^\infty} ,\ea\ee
 \be \la{ww5}\ba  & \left|\int
(1-\eta_\de(x-y))\na K(x-y){\div u}(y) dy\right|\\&\le C\left(
\int_{\de\le |x-y|\le 1} +\int_{ |x-y|> 1} \right) |\na
K(x-y)||{\div u}(y) |dy \\&\le C \int_\de^1 r^{-3}r^2dr\|{\rm
div}u\|_{L^\infty} +C\left(\int_1^\infty r^{-6}r^2dr\right)^{1/2}
\|{\div u}  \|_{L^2}\\&\le - C  \ln \de \|{\rm div}u\|_{L^\infty} +C
\|{\na u} \|_{L^2}.\ea\ee It follows from (\ref{ww2})-(\ref{ww5})
that

\be\la{ww6} \|\na v\|_{L^\infty}\le
C\left(\de^{(q-3)/q}\|\na^2u\|_{L^q}+(1-\ln \de)\|\div
u\|_{L^\infty}+ \|\na u\|_{L^2}\right).\ee Set
$\de=\min\left\{1,\|\na^2u\|_{L^q}^{-q/(q-3)}\right\}$. Then
(\ref{ww6}) becomes \bnn \|\na v\|_{L^\infty}\le C(q)\left(1+\ln
(e+\|\na^2u\|_{L^q})\|\div u\|_{L^\infty}+ \|\na
u\|_{L^2}\right).\enn Therefore (\ref{ww7}) holds.

\section{\la{se3}A priori estimates}

In this section, we will establish some necessary a priori bounds
for smooth solutions to the Cauchy problem (\ref{a1})  (\ref{h1})
(\ref{h2}) to extend the local classical solution guaranteed by
Lemma \ref{th0}. Thus, let $T>0$ be a fixed time and $(\rho,u)$ be
the smooth solution to (\ref{a1})  (\ref{h1})  (\ref{h2})  on
${\r}^3 \times (0,T]$ in the class (\ref{ba1}) with smooth initial
data $(\n_0,u_0)$ satisfying (\ref{co1})-(\ref{co2}). To estimate
this solution, we set $\si(t)\triangleq\min\{1,t \}$ and define
 \bn\la{As1}
  A_1(T) \triangleq \sup_{t\in [0,T] }\left(\sigma\|\nabla u\|_{L^2}^2\right) + \int_0^{T}\int
  \sigma\rho|\dot{u}|^2dxdt,
  \en
   \bn \la{As2}
  A_2(T)  \triangleq\sup_{t\in [0,T] }\sigma^3\int\rho|\dot{u}|^2dx + \int_0^{T}\int
  \sigma^3|\nabla\dot{u}|^2dxdt,
\en
 and
\bnn
  A_3(T)  \triangleq\sup\limits_{0\le t\le T }\int\rho|u|^3(x,t)dx     .
\enn

We have the following key a priori estimates on $(\n,u)$.
\begin{pro}\la{pr1}  Under  the conditions of Theorem
\ref{th1},
  for  \be\la{cc1}{\de_0}\triangleq
(2\beta-1)/(4\beta)\in (0,1/4] ,\ee
   there exists some  positive constant  $\ve$
    depending    on  $\mu$, $\lambda$, $\tn$, $a$, $\ga$, $\on,$ $\beta$ and $M$  such that if
       $(\rho,u)$  is a smooth solution of
       (\ref{a1}) (\ref{h1}) (\ref{h2})  on $\r^3\times (0,T] $
        satisfying
 \be\la{z1}
 \sup\limits_{
 \r^3\times [0,T]}\rho\le 2\bar{\rho},\quad
     A_1(T) + A_2(T) \le 2C_0^{1/2},
 \quad A_3(\si(T))\le 2C_0^{{\de_0}},
  \ee
     the following estimates hold
        \be\la{z2}
 \sup\limits_{\r^3\times [0,T]}\rho\le 7\bar{\rho}/4, \quad
     A_1(T) + A_2(T) \le  C_0^{1/2},
   \quad  A_3(\si(T))\le  C_0^{{\de_0}} ,
  \ee
   provided $C_0\le \ve.$
\end{pro}

{\it Proof. }Proposition \ref{pr1} is an easy consequence of the
following Lemmas \ref{le4}, \ref{le5} and \ref{le7}.

In the following, we will use the convention that $C$ denotes a
generic positive constant
 depending  on $\mu$, $\lambda$, $\tilde{\rho}$, $a$, $\gamma$,
$\bar{\rho},$  $\beta$ and $M$, and   we write $C(\al)$ to emphasize
that $C$ depends on $\al.$

We start with the following   standard   energy estimate for
$(\n,u)$ and preliminary    $L^2$ bounds for $\nabla u$ and
$\rho\dot{u}$.
\begin{lemma}\la{le2}
 Let $(\rho,u)$ be a smooth solution of
 (\ref{a1}) (\ref{h1}) (\ref{h2}) on $\r^3\times (0,T] $
 with $0\le\n(x,t)\le 2\on.$ Then there is a positive constant
  $C=C(\bar{\rho})$ such that
  \bn \la{a16} \sup_{0\le t\le T}\int\left(
\frac{1}{2}\n|u|^2+G(\n)\right)dx +\ia\int \left(\mu |\na
u|^2+(\lambda+\mu) ({\rm div} u)^2\right)dxdt\le C_0,\en
  \be\la{h14}
  A_1(T) \le CC_0 + C\int_0^{T}\int\sigma|\nabla u|^3dxdt,
  \ee
 and
  \be\la{h15}
    A_2(T)
    \le C C_0 + CA_1(T)  + C\int_0^{T}\int\sigma^3 |\nabla u|^4 dxdt.
   \ee
\end{lemma}

{\it Proof.} Multiplying the first  equation in (\ref{a1}) by
$G'(\n)$ and the second by $u^j$ and integrating, applying the far
field condition $(\ref{h1}),$ one shows easily the energy
inequality (\ref{a16}).

The proof of (\ref{h14}) and (\ref{h15}) is due to Hoff\cite{H3}.
  For $m\ge 0,$ multiplying $(\ref{a1})_2 $ by
$\sigma^m \dot{u}$ and then integrating the resulting equality over
${\r^3} $ lead  to \be\la{m0} \ba  \int \sigma^m \rho|\dot{u}|^2dx &
= \int (-\sigma^m \dot{u}\cdot\nabla P + \mu\sigma^m \triangle
 u\cdot\dot{u} + (\lambda+\mu)
 \sigma^m \nabla\text{div}u\cdot\dot{u})dx \\
& \triangleq \sum_{i=1}^{3}M_i. \ea \ee Using $(\ref{a1})_1$ and
integrating by parts give \be\la{m1} \ba
M_1 = & - \int \sigma^m \dot{u}\cdot\nabla Pdx \\
= & \int ( \sigma^m (\text{div}u)_t(P-P(\tn))
- \sigma^m (u\cdot\nabla u)\cdot\nabla P)dx \\
= & \left(\int\sigma^m \text{div}u(P-P(\tn))dx \right)_t
 -  m \sigma^{m-1}\si'\int \text{div}u(P-P(\tn)) dx \\
& + \int\sigma^m \left( P^{'}\rho(\text{div}u)^2
 -  P(\text{div}u)^2+   P\p_iu^j\p_ju^i\right)dx \\
\le & \left(\int\sigma^m \text{div}u(P-P(\tn))dx \right)_t
 + m\si^{m-1}\si'\|\pp\|_{L^2}\|\na u\|_{L^2}
\\& +C(\on)\|\na u\|_{L^2}^2\\
\le & \left(\int\sigma^m \text{div}u(P-P(\tn))dx \right)_t
+C(\on)\|\na u\|_{L^2}^2 +C(\on)m^2\si^{2(m-1)}\si'C_0  . \ea \ee
Integration by parts  implies \be\la{m2} \ba
M_2 & =  \int \mu\sigma^m \triangle u\cdot\dot{u}dx \\
& = -\frac{\mu }{2}\left(\sigma^m\|\nabla u\|_{L^2}^2\right)_t +
\frac{\mu m}{2}\si^{m-1}\si' \|\nabla u\|_{L^2}^2
-\mu \sigma^m \int \p_iu^j\p_i(u^k\p_ku^j)dx \\
& \le  -\frac{\mu }{2}\left(\sigma^m\|\nabla u\|_{L^2}^2\right)_t +
Cm\si^{m-1}\|\na u\|_{L^2}^2 + C \int\sigma^m |\nabla u|^3dx , \ea
\ee and similarly, \be\la{m3}\ba  M_3 &= -
\frac{\lambda+\mu}{2}\left( \sigma^m
\|\text{div}u\|_{L^2}^2\right)_t+  \frac{m(\lambda+\mu)}{2}
\si^{m-1}\|\div u\|_{L^2}^2 \\ &- (\lambda+\mu)\sigma^m \int\div
u\div(u\cdot\na u)dx\\& \le -\frac{\lambda+\mu}{2}\left(\sigma^m
\|\text{div}u\|_{L^2}^2\right)_t+ Cm\si^{m-1}\|\na u\|_{L^2}^2 + C
\int\sigma^m |\nabla u|^3dx  . \ea \ee

Combining (\ref{m0})-(\ref{m3}) leads to \be\la{n1}\ba&
(\si^mB(t))'+\int \sigma^m \rho|\dot{u}|^2dx\\&\le
(Cm\si^{m-1}+C(\on))\|\na u\|_{L^2}^2
 +C(\on)m^2\si^{2(m-1)}\si'C_0 + C
\int\sigma^m |\nabla u|^3dx,\ea\ee where \be \la{n2}\ba
B(t)&\triangleq \frac{\mu  }{2}\|\nabla
u\|_{L^2}^2+\frac{(\lambda+\mu)
}{2}\|\text{div}u\|_{L^2}^2-\int  \text{div}u(P-P(\tn))dx\\
&\ge \frac{\mu }{2}\|\nabla u\|_{L^2}^2+\frac{(\lambda+\mu)
}{2}\|\text{div}u\|_{L^2}^2-C C_0^{1/2}\|\div u\|_{L^2}\\
&\ge \frac{\mu }{4}\|\nabla u\|_{L^2}^2+\frac{(\lambda+\mu)
}{2}\|\text{div}u\|_{L^2}^2-C C_0   .  \ea\ee Integrating (\ref{n1})
over $(0,T),$ choosing $m=1,$ and using (\ref{n2}), one gets
(\ref{h14}).

Next, for   $m \ge 0,$ operating $ \si^m\dot u^j[\pa/\pa t+\div
(u\cdot)]$ to $ (\ref{a1})_2^j,$ summing with respect
 to $j,$ and integrating the resulting equation over ${\r^3}$, one obtains
after integration by parts \be\la{m4} \ba &
\left(\frac{\sigma^m}{2}\int\rho|\dot{u}|^2dx \right)_t
-\frac{m}{2}\sigma^{m-1}\si'\int\rho|\dot{u}|^2dx\\
& =   -\int  \sigma^m \dot{u}^j[\p_jP_t +\text{div}(\p_jPu)]dx +
\mu\int\sigma^m\dot{u}^j[\triangle u_t^j +
\text{div}(u\triangle u^j)] dx\\
&\quad  + (\lambda+\mu)\int\sigma^m\dot{u}^j[\p_t\p_j
 \text{div}u  +\text{div}(u\p_j \text{div}u )] dx \\
& \triangleq\sum_{i=1}^{3}N_i. \ea \ee  It follows from
integration by parts and using the equation $(\ref{a1})_1$ that
\be\la{m5} \ba
N_1 & = - \int \sigma^m\dot{u}^j[\p_jP_t + \text{div}(\p_jPu)]dx \\
& =  \int \sigma^m[-P^{'}\rho\text{div}u\p_j\dot{u}^j +
\p_k(\p_j\dot{u}^ju^k)P - P\p_j(\p_k\dot{u}^ju^k)]dx
\\
&\le C(\on)   \si^m\|\nabla u\|_{L^2}  \|\nabla \dot u\|_{L^2}\\
&\le \de\si^m \|\nabla \dot u\|^2_{L^2} +C(\on,\de)   \si^m\|\nabla
u\|^2_{L^2} . \ea \ee Integration by parts leads to \be\la{m6} \ba
N_2 & =  \mu\int \sigma^m\dot{u}^j[\triangle u_t^j
+ \text{div}(u\triangle u^j)]dx \\
& = - \mu\int \sigma^m[|\nabla\dot{u}|^2 + \p_i\dot{u}^j\p_ku^k\p_iu^j - \p_i\dot{u}^j\p_iu^k\p_ku^j - \p_iu^j\p_iu^k\p_k\dot{u}^j]dx \\
&\le -\frac{ 3\mu}{4} \int \sigma^m|\nabla\dot{u}|^2dx  + C \int
\sigma^m|\nabla u|^4dx . \ea \ee Similarly, \be\la{m7}
 N_3  \le -\frac{\mu+\lambda}{2}
 \int  \sigma^m({\rm div} \dot u)^2dx
  + C \int \sigma^m|\nabla u|^4dx .
\ee Substituting (\ref{m5})-(\ref{m7}) into (\ref{m4}) shows that
for $\de$ suitably small, it holds that\be\la{mm4} \ba & \left(
{\sigma^m} \int\rho|\dot{u}|^2dx \right)_t + { \mu} \int
\sigma^m|\nabla\dot{u}|^2dx+ (\mu+\lambda)
 \int  \sigma^m({\rm div} \dot u)^2dx\\&
\le  {m} \sigma^{m-1}\si'\int\rho|\dot{u}|^2dx+C\si^m\|\na
u\|_{L^4}^4 +C(\on)\si^m\|\na u\|_{L^2}^2.\ea\ee Taking $m=3 $ in
(\ref{mm4}) and noticing that \bnn {3}
\int_0^T\sigma^{2}\si'\int\rho|\dot{u}|^2dxdt\le CA_1(T),\enn
  we immediately obtain (\ref{h15}) after integrating (\ref{mm4})
   over $(0,T).$ The proof of Lemma \ref{le2} is completed.

Next,   the following lemma will play  important roles in  the
estimates on   both $A_i(\si(T)) $ $(i=1, 3)  $ and the uniform
upper bound of the density for small time.

\begin{lemma}\la{zc1} Let $(\rho,u)$ be a smooth solution of (\ref{a1}) (\ref{h1})
(\ref{h2}) on $\r^3\times (0,T] $ satisfying (\ref{z1}). Then there
exist positive constants $K $ and $\ve_0 $ both depending only on
$\mu$, $\lambda$, $\tn$, $a$, $\ga$, $\on,$ $\beta$ and $M$ such
that
   \be\la{uv1}  \sup_{0\le t\le \si(T)}t^{1-\beta}\|\na
u\|_{L^2}^2+\int_0^{\si(T)}t^{1-\beta}\int\n|\dot u|^2dxdt\le
K(\on,M), \ee \be \la{uv2}\sup_{0\le t\le \si(T)}
t^{2-\beta}\int\n|\dot u|^2dx
   +\int_0^{ \si(T)}t^{2-\beta}\int|\na\dot u|^2dxdt\le K(\on,M),\ee
  provided    $C_0\le \ve_0.$

\end{lemma}

{\it Proof.} As in \cite{hof2002}, we define $w_1$ and $w_2$ to be
the solution to: \be \la{sas2}Lw_1 =0 ,\quad w_1(x,0)=w_{10}(x), \ee
and\be\la{sas3} Lw_2 =-\nabla P(\rho),\quad w_2(x,0)=0,\ee
respectively, with  $L$ being the linear differential operator
defined by \be\ba(Lw)^j&\triangleq \rho w^j_{t}+ \rho u\cdot\na w^j
-(\mu\Delta w^j+(\mu+\lambda) \div w_{x_j})\no&=\n\dot w^j
-(\mu\Delta w^j+(\mu+\lambda) \div w_{x_j}),\quad j=1,2,3. \ea\ee

Straightforward energy estimates  show  that: \be\la{sas4}
\sup_{0\le t\le\si(T)}\int \rho|w_1|^2dx+\int^{\si(T)}_{0}
\int|\nabla w_1|^2dxdt\leq C(\overline{\rho})\int |w_{10}|^2dx,\ee
and \be\la{sas1} \sup_{0\le t\le\si(T)}\int
\rho|w_2|^2dx+\int^{\si(T)}_{0} \int|\nabla w_2|^2dxdt\leq
C(\overline{\rho})C_0 . \ee It follows from (\ref{sas2}) and
standard $L^2$-estimate for elliptic  system that \be\la{sas5}\|\na
w_1\|_{L^6}\le C\|\na^2w_1\|_{L^2}\le C\|\n\dot w_1\|_{L^2}. \ee

 Multiplying (\ref{sas2}) by
$ w_{1t} $ and integrating the resulting equality over $\O,$ we get
by (\ref{sas5}) and $(\ref{z1})_3$ that
 \be\ba& \frac{1}{2}\left( \mu\|\na {w_1}\|_{L^2}^2
 +(\mu+\lambda)\|\div {w_1}\|^2_{L^2} \right)_t
 +\int\n|\dot {w_1}|^2dx\no&=\int \n\dot {w_1}(u\cdot\na {w_1})dx
  \\&\le  C(\on)\left(\int\n|\dot {w_1}|^2dx\right)^{1/2}
  \left(\int\n|u|^3dx\right)^{1/3}\|\na {w_1}\|_{L^6}
  \\&\le C(\on)C_0^{{\de_0}/3}\int\n|\dot {w_1}|^2dx,
  \ea\ee
which, together with Gronwall's inequality  and (\ref{sas4}), gives
\bn \label{uu1} \sup_{0\le t\le \si(T)}\|\nabla
w_1\|_{L^2}^2+\int_0^{\si(T)}\int\n|\dot {w_1}|^2dxdt\leq C\|\na
w_{10}\|_{L^2}^2,\en and \bn  \label{uu2} \sup_{0\le t\le
\si(T)}t\|\nabla w_1\|_{L^2}^2+\int_0^{\si(T)}t\int\n|\dot
{w_1}|^2dxdt\leq C\| w_{10}\|_{L^2}^2,\en provided $C_0\le
\ve_{01}\triangleq (2C(\on))^{-3/{\de_0}}.$

Since the solution operator $w_{10}\mapsto w_1(\cdot,t)$ is linear,
by the   standard Stein-Weiss interpolation argument (\cite{bl}),
one can deduce from (\ref{uu1}) and (\ref{uu2}) that for any
$\theta\in [\beta,1],$\bn \label{uu4} \sup_{0\le t\le
\si(T)}t^{1-\theta}\|\nabla
w_1\|_{L^2}^2+\int_0^{\si(T)}t^{1-\theta}\int\n|\dot
{w_1}|^2dxdt\leq C\| w_{10}\|_{\dot H^\theta}^2,\en with a uniform
constant $C$ independent of $\theta.$

Next, we estimate $w_2.$ It follows from a similar way to
(\ref{h19}) and (\ref{h18}) that \be\la{sas6}
\begin{cases}\|\na((2\mu+\lambda)\div w_2-(\pp))\|_{L^2}\le C\|\n \dot
w_2\|_{L^2},\\ \|\na w_2\|_{L^6}\le C(\|\n\dot
w_2\|_{L^2}+\|\pp\|_{L^6}).\end{cases}\ee  Multiplying (\ref{sas3})
by $w_{2t} ,$  integrating the resultant equation over $\O $ and
using (\ref{sas6}), one has
 \be\ba& \frac{1}{2}\left( \mu\|\na {w_2}\|_{L^2}^2
 +(\mu+\lambda)\|\div {w_2}\|^2_{L^2} -2\int(\pp)\div w_2dx \right)_t
 +\int\n|\dot {w_2}|^2dx\no&=\int \n\dot {w_2}(u\cdot\na
 {w_2})dx-\int P_t\div w_2dx
  \\&\le  C(\on)\left(\int\n|\dot {w_2}|^2dx\right)^{1/2}
  \left(\int\n|u|^3dx\right)^{1/3}\|\na {w_2}\|_{L^6}\\&\quad+
\int\div w_2\div ((\pp)u)dx+\int (\tp+(\ga-1)P) \div u \div w_2 dx
 \\&\le  C(\on)C_0^{{\de_0}/3}\left(\int\n|\dot {w_2}|^2dx\right)^{1/2}
\left(\|\n^{1/2} \dot w_2\|_{L^2}+\|\pp\|_{L^6}\right)\\&\quad
-\int(\pp)u\cdot\na \left(\div
w_2-\frac{\pp}{2\mu+\lambda}\right)dx\\&\quad+\frac{1}{2(2\mu+\lambda)}
\int(\pp)^2  \div u dx+C\|\na u\|_{L^2}^2+C\|\na w_2\|_{L^2}^2
\\&\le  C(\on)C_0^{{\de_0}/3} \int\n|\dot {w_2}|^2dx
 +CC_0^{1/3}+C\|\pp\|_{L^3}\|u\|_{L^6}\|\n^{1/2}\dot w_2\|_{L^2}
 \\&\quad   +C\|\pp\|_{L^4}^4 +C\|\na u\|_{L^2}^2
 +C\|\na w_2\|_{L^2}^2\\&\le  C(\on)C_0^{{\de_0}/3} \int\n|\dot {w_2}|^2dx
 +CC_0^{1/3} +C\|\na u\|_{L^2}^2
 +C\|\na w_2\|_{L^2}^2,
  \ea\ee
 which, together with (\ref{sas1}) and Gronwall's inequality, gives
\bn  \label{uu3} \sup_{0\le t\le \si(T)}\|\nabla
w_2\|_{L^2}^2+\int_0^{\si(T)}\int\n|\dot {w_2}|^2dxdt\leq
CC_0^{1/3},\en   provided $C_0\le \ve_{02}\triangleq
(2C(\on))^{-3/{\de_0}}.$ Taking $w_{10}=u_0 $ so that $w_1+w_2=u,$
we then conclude from (\ref{uu4}) and (\ref{uu3}) that for any
$\theta\in[\beta,1],$
 \be\la{uv5}  \sup_{0\le t\le \si(T)}t^{1-\theta}\|\na
u\|_{L^2}^2+\int_0^{\si(T)}t^{1-\theta}\int\n|\dot u|^2dxdt\le
C\|u_0\|_{\dot H^\theta}^2+CC^{1/3}_0, \ee provided $C_0\le
\ve_0\triangleq \min\{\ve_{01},\ve_{02}\}.$  Thus, (\ref{uv1})
follows from (\ref{uv5}) directly.

To prove (\ref{uv2}), we take $m=2-\beta$ in (\ref{mm4}) to obtain,
after integrating (\ref{mm4}) over $(0,\si(T)) $ and using
(\ref{uv5}) and (\ref{h17}), that \be\ba &\sup_{0\le t\le  \si(T)}
t^{2-\beta}\int\n|\dot u|^2dx
   +\int_0^{ \si(T)}t^{2-\beta}\int|\na\dot u|^2dxdt\\&
   \le  C\int_0^{\si(T)}t^{2-\beta}\|\na u\|_{L^4}^4dt+C(\on,M)\\&
   \le  C\int_0^{\si(T)}t^{2-\beta}\|\na u\|_{L^2}
   \left(\|\n\dot u\|_{L^2}^3+\|\pp\|_{L^6}^3\right)  dt+C(\on,M)\\&
   \le C\int_0^{\si(T)}t^{(2\beta-1)/2}\left( t^{ 1-\beta }\|\na
   u\|^2_{L^2}\right)^{1/2}(t^{ 2-\beta }\|\n^{1/2}\dot u\|^2_{L^2})^{1/2}
    (t^{1-\beta}\|\n^{1/2}\dot u\|_{L^2}^2 ) dt\no &\quad+ C(\on,M) \\&\le
  C(\on,M) \left(\sup_{0\le t\le  \si(T)}
t^{2-\beta}\int\n |\dot u|^2dx\right)^{1/2}+ C(\on,M),\ea\ee which
implies (\ref{uv2}). Thus, we finish the proof of Lemma \ref{zc1}.

The following Lemma \ref{le4} will give an estimate on
$A_3(\si(T)).$
\begin{lemma}\la{le4}   If  $(\rho,u)$ is a smooth solution  of
   (\ref{a1}) (\ref{h1}) (\ref{h2})
    on $\r^3\times (0,T] $ satisfying (\ref{z1}),
    there exists a positive constant $\ve_1$ depending on $\mu$, $\lambda$, $\tn$, $a$, $\ga$, $\on,$ $\beta$ and $M$ such
that the following estimate holds for ${\de_0}$ defined by
(\ref{cc1}):
  \be\la{z200}
  \sup_{0\le t\le \si(T)}\int\rho|u|^3(x,t)dx \le
  C_0^{{\de_0} },
  \ee
provided $C_0\le \ve_1.$

\end{lemma}

{\it Proof.} Multiplying $(\ref{a1})_2$ by $3|u| u$, and integrating
the resulting equation over $\O$, we obtain by (\ref{h17})  that \be
\ba& \frac{d}{dt}\int \rho|u|^3dx
 \\& \le C\int  |u| |\nabla u|^2dx+C\int |P-P(\tn)|
 |u| |\na u| dx \\ &\le C\|u\|_{L^6} \|\na
u\|_{L^2}^{3/2}\|\na u\|_{L^6}^{1/2}+C
\|P-P(\tn)\|_{L^{3 }}\|u\|_{L^6} \|\na u\|_{L^2} \\
&\le C \|\na u\|_{L^2}^{5/2}\left( \|\n \dot
u\|_{L^2}+\|P-P(\tn)\|_{L^6} \right)^{1/2} +CC_0^{1/6}\|\na
u\|_{L^2}^2\\ &\le C \|\na u\|_{L^2}^{5/2}\left(\|\n \dot u\|_{L^2}
+C_0^{1/6}\right)^{1/2}+CC_0^{1/6}\|\na u\|_{L^2}^2\\
&\le Ct^{(2{\de_0}-3/2)(1-\beta)  }(t^{1-\beta}\|\na
u\|_{L^2}^{2})^{-2{\de_0}+5/4}(t^{1-\beta}\|\n^{1/2}\dot
u\|_{L^2}^2)^{1/4}\|\na u\|_{L^2}^{4{\de_0}}\no&\quad
+CC_0^{1/12}t^{-3(1-\beta)/4}(t^{1-\beta}\|\na u\|_{L^2}^{2})^{3/4}
\|\na u\|_{L^2}+CC_0^{1/6}\|\na u\|_{L^2}^2,\ea \ee which together
with (\ref{uv1}) and  (\ref{a16}) gives \be\la{zp201}\ba& \sup_{0\le
t\le \si(T)}\int\n|u|^3dx\\&\le
 C(\on,M)\left(\int_0^{\si(T)}
t^{-\frac{2(3-4{\de_0})(1-\beta)}{3-8{\de_0}}}dt\right)^{(3-8{\de_0})/4}
\left(\int_0^{\si(T)}\|\na u\|_{L^2}^2dt\right)^{2{\de_0}}\\&\quad+C
(\on,M)C_0^{1/12} \left(\int_0^{\si(T)}
t^{-3(1-\beta)/2}dt\right)^{1/2} \left(\int_0^{\si(T)}\|\na
u\|_{L^2}^2dt\right)^{1/2}\\&\quad+\int\n_0|u_0|^3dx+CC_0
 \\&\le C(\on,M)C_0^{2{\de_0} }, \ea\ee provided $C_0\le\ve_0,$ where in the last inequality we
 have used
the following simple facts: \be\la{uv4}\ba\int\n_0|u_0|^3dx&\le
C\left(\int\n_0|u_0|^2dx\right)^{3(2\beta-1)/(4\beta)} \|u_0\|_{\dot
H^\beta}^{3/(2\beta)} \\&\le C(\on,M)C_0^{2{\de_0}} ,\ea\ee and
\bnn\frac{2(3-4{\de_0})(1-\beta)}{3-8{\de_0}}
=1-\frac{\beta(2\beta-1)}{2-\beta}<1 \enn due to (\ref{cc1}) and
$\beta\in (1/2,1].$ Thus, it follows from (\ref{zp201}) that
(\ref{z200}) holds provided $C_0\le \ve_1,$ where\bnn\ve_1\triangleq
\min\left\{\ve_0,(C(\on,M))^{-1/\de_0}\right\}=
\min\left\{\ve_0,(C(\on,M))^{-4\beta/{(2\beta-1)}}\right\}.\enn The
proof of Lemma \ref{le4} is completed.

\begin{lemma}\la{le5}
      There exists a positive constant
    $\ve_2(  \mu,\lambda,\tn, a,\ga,\on,\beta, M  )\le \ve_1$ such that,  if  $(\rho,u)$ is a smooth solution  of
   (\ref{a1}) (\ref{h1}) (\ref{h2})     on $\r^3\times (0,T] $ satisfying (\ref{z1}), then
  \be\la{h27}
  A_1(T)+A_2(T)\le C_0^{1/2},
  \ee provided $C_0\le \ve_2. $
   \end{lemma}
{\it Proof.}  Lemma \ref{le2} shows that   \be\la{h28}
A_1(T)+A_2(T)\le C(\on)C_0 + C(\on)\int_0^{T}     \sigma^3\|\nabla
u\|_{L^4}^4 ds+C(\on)\int_0^{T} \sigma\|\nabla u\|_{L^3}^3ds. \ee
Due to (\ref{h18}), \be\la{h29} \int_0^{T} \sigma^3\|\nabla
u\|_{L^4}^4 ds\le C\int_0^{T}\sigma^3 \left(\|F\|_{L^4}^4 +
\|\o\|_{L^4}^4\right) ds + C\int_0^{T}\sigma^3\|P-P(\tn)\|_{L^4}^4
ds. \ee It follows from (\ref{h20}) that \be\la{h99} \ba
&   \int_0^{T}\sigma^3 \left(\|F\|_{L^4}^4 + \|\o\|_{L^4}^4\right) ds\\
& \le C\int_0^{T}\sigma^3\left(\|\nabla u\|_{L^2}+\|\pp\|_{L^2}\right)\|\n \dot u\|_{L^2}^3ds\\
& \le C(\on) \sup_{t\in (0,T]}\left( \si^{3/2}\|\sqrt{\n} \dot
u\|_{L^2}  \left(\si^{1/2}\|\nabla u\|_{L^2}
+C_0^{1/2}\right)\right)
\int_0^{T}\int\sigma\rho|\dot{u}|^2dxds\\
&\le C(\on)\left(A_1^{1/2}(T)+C_0^{1/2}\right)A_2^{1/2}(T)A_1(T)\\
&\le C(\on)C_0. \ea \ee

To estimate the second term on the right hand side of (\ref{h29}),
one deduces from $(\ref{a1})_1$ that $\pp$ satisfies \be
\la{a95}(\pp)_t+u\cdot\nabla (\pp)+\ga(\pp){\rm div}u+\ga P(\tn){\rm
div}u=0.\ee Multiplying (\ref{a95}) by $3 (\pp)^2$ and integrating
the resulting equality over ${\r^3} ,$ one gets after using ${\rm
div}u=\frac{1}{2\mu+\lambda}(F+\pp)$ that \bn
\la{a96}\lefteqn{\frac{3\ga-1}{2\mu+\lambda}\|\pp\|_{L^4}^4 }\no&&
=-\left(\int(\pp)^3dx\right)_t-\frac{3\ga-1}{2\mu+\lambda}\int
(\pp)^3Fdx\no&&\quad-3\ga P(\tn)\int (\pp)^2{\rm div}udx\no&&
\le-\left(\int(\pp)^3dx\right)_t+\eta
\|\pp\|_{L^4}^4+C_\eta\|F\|_{L^4}^4+C_\eta\|\nabla u\|_{L^2}^2.\en
Multiplying (\ref{a96}) by $\si^3$, integrating the resulting
inequality over $(0,T),$ and choosing $\eta$ suitably small, one may
arrive at \be\la{h30} \ba
&\int_0^{T} \sigma^3\|\pp\|_{L^4}^4 dt \\
& \le C\sup_{0\le t\le T}\|P-P(\tn)\|^3_{L^3}+C\int_0^{\si(T)}
\|P-P(\tn)\|^3_{L^3}dt\\&+C(\on)\int_0^{T} \sigma^3\|F\|^4_{L^4} ds +C(\on) C_0 \\
& \le  C(\on)C_0, \ea \ee where (\ref{h99}) has been used.
Therefore, collecting (\ref{h29}), (\ref{h99}) and (\ref{h30})
shows that \be\la{j29} \int_0^{T} \sigma^3\left(\|\nabla
u\|_{L^4}^4+\|\pp\|_{L^4}^4\right) ds\le C(\on)C_0.\ee

Finally, we estimate the last term on the right hand side of
(\ref{h28}). First, (\ref{j29}) implies that \be\la{h33} \ba &
\int_{\si(T)}^{T}\int\sigma|\nabla u|^3dxds \le
\int_{\si(T)}^{T}\int( |\nabla u|^4 + |\nabla u|^2)dxds  \le  C C_0.
\ea \ee Next, one deduces from (\ref{h17}), (\ref{uv1}) and
(\ref{z1})  that \be\la{h34} \ba
& \int_0^{\si(T)} \sigma\|\nabla u\|_{L^3}^3 dt\\
& \le  C(\on)\int_0^{\si(T)}t  \|\nabla u\|_{L^2}^{3/2} \left(\|\rho
\dot{u}\|^{3/2}_{L^2}+C_0^{1/4}\right)dt
\\& \le C(\on)\int_0^{\si(T)}\left( t^{(1-\beta)/2}\|\nabla
u\|_{L^2}    \right)\|\nabla u\|_{L^2}^{1/2}   \left(t
\int\rho|\dot{u}|^2dx\right)^{3/4}dt+ C(\on)C_0
\\  & \le C(\on)\sup_{t\in (0,\si(T)]} \left(t^{(1-\beta)/2}
\|\nabla u\|_{L^2}\right)     \int_0^{\si(T)}  \|\nabla
u\|_{L^2}^{1/2}
\left(t \int\rho|\dot{u}|^2dx\right)^{3/4}dt\\
&\quad +C(\on)C_0\\&\le C(\bar \n,M )A_1^{3/4}C_0^{1/4}  +
C(\on)C_0\\&\le C(\bar \n,M) C_0^{5/8} , \ea \ee  provided $C_0\le
\ve_1.$ It thus follows from (\ref{h28}) and (\ref{j29})-(\ref{h34})
that the left hand side of (\ref{h27}) is bounded by \bnn\la{h35}
C(\bar \n,M)C_0^{5/8}\le C_0^{1/2} \enn provided \bnn\la{h36} C_0\le
\ve_2\triangleq\min\left\{\ve_1,\left(C(\bar \n,M)\right)^{-8}
\right\}.\enn The proof of Lemma \ref{le5} is completed.

We now proceed to derive a uniform (in time) upper bound for the
density, which turns out to be the key to obtain all the higher
order estimates and thus to extend the classical solution globally.
We will use an approach motivated by our previous study on the
two-dimensional Stokes approximation equations (\cite{lx}).

\begin{lemma}\la{le7}
There exists a positive constant
   $\ve=\ve (\on ,M) $ as described in Theorem \ref{th1} such that,
    if  $(\rho,u)$ is a smooth solution  of
   (\ref{a1}) (\ref{h1}) (\ref{h2})     on $\r^3\times (0,T] $
   satisfying (\ref{z1}), then
      \bnn\sup_{0\le t\le T}\|\n(t)\|_{L^\infty}  \le
\frac{7\bar \n }{4}  ,\enn
      provided $C_0\le \ve . $

   \end{lemma}

{\it Proof.} Rewrite the equation of the mass conservation
$(\ref{a1})_1$ as \bnn D_t \n=g(\n)+b'(t), \enn where \bnn
D_t\n\triangleq\n_t+u \cdot\nabla \n ,\quad
g(\n)\triangleq-\frac{a\n}{2\mu+\lambda}(\n^{\ga }-\tn^\ga) ,
\quad b(t)\triangleq-\frac{1}{2\mu+\lambda} \int_0^t\n Fdt. \enn

For $t\in [0,\si(T)],$ one deduces from Lemma \ref{l1}, (\ref{h19}),
(\ref{h27}), (\ref{uv1}), (\ref{uv2}) and (\ref{g2}) that for
${\de_0}$  as in (\ref{cc1}) and for all $ 0\le t_1<t_2\le \si(T),$
\be\ba &|b(t_2)-b(t_1)|\\&\le C\int_0^{\si(T)}\|(\n
F)(\cdot,t)\|_{L^\infty}dt\\ &\le C(\on)
\int_0^{\si(T)}\|F(\cdot,t)\|^{1/2}_{L^6}\|\na
F(\cdot,t)\|^{1/2}_{L^6}dt\no&\le  C(\on) \int_0^{\si(T)}
\|\n^{1/2}\dot u\|^{1/2}_{L^2} \|\na\dot u\|^{1/2}_{L^2}dt \no&\le
C(\on)\int_0^{\si(T)}t^{-(2-\beta)/4}\|\n\dot
u\|_{L^2}^{1/2}\left(t^{2-\beta}\|\na\dot u\|_{L^2}^2\right)^{1/4}dt
\no&\le C(\on,M)\left(\int_0^{\si(T)}t^{-(2-\beta)/3}\|\n\dot
u\|_{L^2}^{2/3} dt\right)^{3/4} \no&=
C(\on,M)\left(\int_0^{\si(T)}t^{-[(2-\beta)(-\de_0+2/3)+\de_0]}
\left(t^{2-\beta}\|\n^{1/2}\dot
u\|_{L^2}^2\right)^{-\de_0+1/3}\left(t\|\n^{1/2}\dot
u\|_{L^2}^2\right)^{\de_0} dt\right)^{3/4} \no&\le C(\on,M)
(A_1(\si(T)))^{3\de_0/4}\no&\le C(\on,M)C_0^{3\de_0/8},\ea\ee
provided $C_0\le\ve_2.$ Therefore, for $ t\in [0,\si(T)],$ one can
choose $N_0$ and $N_1$ in (\ref{a100}) as follows:\bnn N_1=0, \quad
N_0=C(\on,M)C_0^{3{\de_0}/8}, \enn and $\bar\zeta=\tn$ in
(\ref{a101}). Then
$$ g(\zeta)=-\frac{a\zeta}{2\mu+\lambda}(\zeta^{\ga}-\tn^{\ga})
\le -N_1=0, \quad \mbox{for all}\quad { \zeta}\ge \bar\zeta= \tn.$$
Lemma \ref{le1} thus yields that \be\la{a103}\sup_{t\in
[0,\si(T)]}\|\rho\|_{L^\infty}\le \max\{  \bar\n ,\tn\}+N_0\le \on
+C(\on,M)C_0^{3{\de_0}/8}\le\frac{3 \bar\n  }{2},\ee
 provided $$C_0\le\min\{\ve_2,\ve_3\},\quad\mbox{ for }
 \ve_3\triangleq \left(\frac{\bar \n }{2C(\on,M) }\right)^{8/(3{\de_0})}
 =\left(\frac{\bar \n }{2C(\on,M) }\right)^{32\beta/(3(2\beta-1))}.$$

On the other hand, for $t\in [\si(T),T]$, one deduces from Lemma
\ref{l1}, (\ref{h27}), (\ref{a16}),
 and (\ref{h19})
that for all $\si(T)\le t_1\le t_2\le T,$ \be\ba  |b(t_2)-b(t_1)|
 &\le  C(\on) \int_{t_1}^{t_2}\|F(\cdot,t)\|_{L^\infty}dt \no& \le
\frac{a}{2\mu+\lambda}(t_2-t_1)+C(\on)
\int_{\si(T)}^{T}\|F(\cdot,t)\|^{8/3}_{L^\infty}dt\no & \le \frac{
a}{2\mu+\lambda}(t_2-t_1)+C(\on)
\int_{\si(T)}^T\|F(\cdot,t)\|^{2/3}_{L^2}\|\na
F(\cdot,t)\|^{2}_{L^6}dt \no& \le \frac{a}{2\mu+\lambda}(t_2-t_1)+ C
(\bar\n)C_0^{1/6}\int_{\si(T)}^T \|\na \dot
u(\cdot,t)\|^{2}_{L^2}dt\no& \le \frac{a}{2\mu+\lambda}(t_2-t_1)+ C
(\bar\n)C_0^{2/3},\ea\ee provided $C_0\le \ve_2.$ Therefore, one can
choose $N_1$ and $N_0$ in (\ref{a100}) as: \bnn
N_1=\frac{a}{2\mu+\lambda}, \quad N_0=C(\on)C_0^{2/3}.\enn Note that
$$ g(\zeta)=-\frac{a\zeta}{2\mu+\lambda}(\zeta^{\ga}-\tn^\ga)\le -N_1
=-\frac{ a}{2\mu+\lambda}, \quad \mbox{for all}\quad { \zeta}\ge
\tn+1.$$ So one can set $\bar\zeta=\tn+1 $ in (\ref{a101}). Lemma
\ref{le1} and (\ref{a103}) thus yield that \be\la{a102} \sup_{t\in
[\si(T),T]}\|\rho\|_{L^\infty}\le \max\left\{ \frac{ 3\bar \n
}{2},\tn+1\right\}+N_0\le \frac{3\bar \n }{2 } +C(\on)C^{2/3}_0\le
\frac{7\bar \n }{4} ,\ee provided \be\la{g6} C_0\le
\ve\triangleq\min\{\ve_2,\ve_3,\ve_4\}, \quad\mbox{ for
}\ve_4\triangleq \left(\frac{ \bar \n }{4C(\on) }\right)^{3/2}.\ee
The combination of (\ref{a103}) with (\ref{a102}) completes the
proof of Lemma \ref{le7}.

 From now on, we will always assume
that the initial energy $C_0$ satisfies  (\ref{g6}) and the positive
constant $C $ may depend on \bnn  T,\,\, \|\n_0^{1/2}g\|_{L^2},\,\,
\|\na g\|_{L^2} , \,\,\|\na u_0\|_{H^2},\,\,
    \|\n_0-\tn\|_{H^3}  ,  \,\, \|P(\n_0)-P(\tn)\|_{H^3} , \,\,\enn
besides $\mu$, $\lambda$, $\tn$, $a$, $\ga$, $\on,$  $\beta$ and $
M,$ where $g$ is as in (\ref{co2}).

Next, we will derive important estimates on the spatial gradient of
the smooth solution $(\rho,u)$.

\begin{lemma}\la{le11}
 The following estimates hold
  \be\label{lee2}
\sup_{0\le t\le T}\int\rho|\dot u|^2dx + \ia\int|\nabla\dot
u|^2dxdt\le C,\ee
 \be\la{qq1}
  \sup_{0\le t\le T}\left(\norm[L^2\cap L^6]{\nabla\rho}
   + \|\nabla u\|_{H^1}
  \right) + \int_0^{T} \norm[L^{\infty}]{\nabla u}
   dt\le C .
  \ee

\end{lemma}

{\it Proof. } Taking $\theta=1$ in (\ref{uv5}) together with
(\ref{h27}) gives
     \be\la{a93}
  \sup_{t\in[0,T]}\|\nabla u\|_{L^2}^2 + \int_0^{T}\int\rho|\dot{u}|^2dxdt
  \le C .
  \ee
Taking $m=0$ in (\ref{mm4}), one can deduce from
Gagliardo-Nirenberg's inequality (\ref{g1}), (\ref{h19}),
(\ref{a93}) and (\ref{mm4}) that\be\la{mm5} \ba & \left(
\int\rho|\dot{u}|^2dx \right)_t + { \mu} \int |\nabla\dot{u}|^2dx+
(\mu+\lambda)
 \int  ({\rm div} \dot u)^2dx\\&
\le C \|\na u\|_{L^4}^4 +C(\on)  \|\na u\|_{L^2}^2\\& \le C \|\na
u\|_{L^2} \|\na u\|_{L^6}^3 +C\\& \le C  \left(
\|F\|_{L^6}^3+\|\o\|_{L^6}^3+\|P-P(\tn)\|_{L^6}^3\right) +C\\& \le C
\left( \|\na F\|_{L^2}^3+\|\na \o\|_{L^2}^3 \right) +C \\& \le C
 \|\n \dot u\|_{L^2}^3  +C\\& \le C
 \|\n^{1/2} \dot u\|_{L^2}^4  +C .\ea\ee
Taking into account on the compatibility condition (\ref{co2}), we
can define \be \la{mm6}\sqrt{\n} \dot u(x,t=0)=\sqrt{\n_0}g.\ee Then
(\ref{lee2}) follows from (\ref{a93})-(\ref{mm6}) and Gronwall's
inequality.

Next, we prove (\ref{qq1}) by using Lemma \ref{le9} as in
\cite{hlx}. For $ 2\le p\le 6,$ $|\nabla\rho|^p$ satisfies \bnnn \ba
& (|\nabla\rho|^p)_t + \text{div}(|\nabla\rho|^pu)+ (p-1)|\nabla\rho|^p\text{div}u  \\
 &+ p|\nabla\rho|^{p-2}(\nabla\rho)^t \nabla u (\nabla\rho) +
p\rho|\nabla\rho|^{p-2}\nabla\rho\cdot\nabla\text{div}u = 0.\ea
\ennn Thus, \be\la{L11}\ba
\partial_t\norm[L^p]{\nabla\rho}&\le
 C(1+\norm[L^{\infty}]{\nabla u} )
\norm[L^p]{\nabla\rho} +C\|\na^2u\|_{L^p}\\ &\le
 C(1+\norm[L^{\infty}]{\nabla u} )
\norm[L^p]{\nabla\rho} +C\|\n\dot u\|_{L^p}, \ea\ee due to\be
\la{ua1}\|\na^2 u\|_{L^p}\le   C\left(\|\n\dot u\|_{L^p}+ \|\nabla
P \|_{L^p}\right),\ee which follows from the standard
$L^p$-estimate for the following elliptic system:
 \bn\la{zp202}  \mu\Delta
u+(\mu+\lambda)\na {\rm div}u=\n \dot u+\na P,\quad \, u\rightarrow
0\,\,\mbox{ as } |x|\rightarrow \infty. \en  It follows from Lemma
\ref{le9} and (\ref{ua1}) that
    \be\la{u13}\ba  \|\na
u\|_{L^\infty } &\le C\left(\|{\rm div}u\|_{L^\infty }+
\|\o\|_{L^\infty } \right)\log(e+\|\na^2 u\|_{L^6 }) +C\|\na
u\|_{L^2} +C \\&\le C\left(\|{\rm div}u\|_{L^\infty }+
\|\o\|_{L^\infty } \right)\log(e+\|\dot u\|_{L^6 } +\|\na P\|_{L^6})
+C \\&\le C\left(\|{\rm div}u\|_{L^\infty }+ \|\o\|_{L^\infty }
\right)\log(e+\|\na\dot u\|_{L^2 } )\\&\quad+C\left(\|{\rm
div}u\|_{L^\infty }+ \|\o\|_{L^\infty } \right)\log(e  +\|\na
\n\|_{L^6}) +C. \ea\ee  Set \bnn f(t)\triangleq e+\|\na
\n\|_{L^6},\quad g(t)\triangleq 1+ \left(\|{\rm div}u\|_{L^\infty }+
\|\o\|_{L^\infty } \right)\log(e+\|\na \dot u\|_{L^2})+\|\na \dot
u\|_{L^2}.\enn Combining (\ref{u13}) with (\ref{L11}) and setting
$p=6$ in
 (\ref{L11}), one gets
\bnn f'(t)\le C g(t) f(t)+ C g(t) f(t)\ln f(t)+Cg(t), \enn which
yields \be\la{hb1} (\ln f(t))'\le Cg(t)+Cg(t)\ln f(t),\ee due to
$f(t)>1.$ Note that (\ref{hj1}), Lemma \ref{l1}, (\ref{h19}),
(\ref{lee2}), and Lemma \ref{le7} imply
 \be \ba\la{hb2}  \int_0^Tg(t)dt &\le  C \int_0^T\left(\|{\rm
div}u\|^2_{L^\infty}+\|\o\|^2_{L^\infty} \right)dt+C
 \\&\le C\int_0^T\left(\|F\|^2_{L^\infty}+\|P-P(\tn)\|^2_{L^\infty}
 +\|\o\|^2_{L^\infty}\right)dt+C\\&\le
C\ia\left(\| F\|^2_{L^2}+\|\nabla F\|^2_{L^6}+\| \o\|^2_{
L^2}+\|\nabla \o\|^2_{  L^6}\right)dt+C    \\ &\le C\ia \|\nabla
\dot
u\|_{L^2}^{2} dt +C\\
&\le C,\ea\ee
 which, together with (\ref{hb1}) and Gronwall's inequality, shows
that \bnn \sup\limits_{0\le t\le T}
   f(t)\le C.\enn
Consequently, \bn \la{u113} \sup\limits_{0\le t\le T}\|\nabla
\rho\|_{L^6}\le C.\en  As a consequence of (\ref{u13}), (\ref{hb2})
and (\ref{u113}), one obtains
  \be \la{v6}\ia\|\nabla u\|_{L^\infty}dt\le C.\ee  Next,
taking $p=2$ in (\ref{L11}), one gets by using (\ref{v6}),
(\ref{a93}) and Gronwall's inequality that \bnn \sup\limits_{0\le
t\le T}\|\nabla \rho\|_{L^2}\le C,\enn which, together with
(\ref{ua1}), (\ref{lee2}), (\ref{a93}), (\ref{u113}), and
(\ref{v6}), gives (\ref{qq1}). The proof of Lemma \ref{le11} is
completed.

The following Lemmas \ref{le9-1}-\ref{sq90} will deal with the
higher order estimates of the solutions which are needed to
guarantee the extension of local classical solution to be a global
one. The proofs are similar to the ones in \cite{hx2}, and we sketch
them here for completeness.

\begin{lemma}\la{le9-1}
 The following estimates  hold \be\label{ja1}
\sup_{0\le t\le T}\int\rho|  u_t|^2dx + \ia\int|\nabla u_t|^2dxdt\le
C,\ee \be\la{va1} \sup_{t\in[0,T]}(\norm[H^2]{\rho-\tilde{\rho}} +
\norm[H^2]{P(\rho)-P(\tilde{\rho})}) \le C.\ee
\end{lemma}

{\it Proof. } Estimate (\ref{ja1}) follows directly from the
following simple facts:\be\ba \int\rho|
u_t|^2dx &\le \int\rho| \dot u |^2dx+\int \n |u\cdot\na u|^2dx\\
&\le C+C\|\n^{1/2}u\|_{L^2}\|u\|_{L^6}\|\na u\|^2_{L^6}\no &\le
C,\ea \ee and \be \la{ja2}\ba
 \|\nabla u_t\|_{L^2}^2 &\le   \|\nabla \dot
u\|_{L^2}^2+\|\nabla(u\cdot\nabla u)\|_{L^2}^2  \no &\le\|\nabla
\dot u\|_{L^2}^2+C  \|u\|_{L^\infty}^2\|\nabla^2
  u \|_{L^2}^2+ C\| \nabla u \|_{L^4}^4 \\ &\le \|\nabla \dot
u\|_{L^2}^2+
 C,\ea\ee due to Lemma  \ref{le11}.

 Next, we prove (\ref{va1}). Note that $P $
 satisfies \bn \la{p0} P_t+u\cdot\nabla P+\ga P{\rm
div}u=0,\en which, together with $(\ref{a1})_1$ and a simple
computation, yields that\be\la{ua2}
 \ba\lefteqn{
\frac{d}{dt}\left(\norm[L^2]{\nabla^2P }^2 +\norm[L^2]{\nabla^2 \rho
}^2\right)}\\& & \le C(1+\norm[L^{\infty}]{\nabla
u})\left(\norm[L^2]{\nabla^2P }^2 +\norm[L^2]{\nabla^2 \rho
}^2\right) + C\norm[H^2]{F}^2 +C\norm[H^2]{\o}^2 +C, \ea\ee
 where we have used the following simple fact:
\be\ba \la{va2}\|\nabla u\|_{H^m}&\le C\left( \|{\rm div}u\|_{H^m}+
\| \o\|_{H^m}  \right)\no&\le C\left( \|F\|_{H^m}+ \|\o\|_{H^m}
+\|P-P(\tn)\|_{H^m}\right),\quad \mbox{for }\,\,m=1,2 .\ea\ee
 Noticing that $F$ and $\o$
satisfy (\ref{h13}),
 we get by the standard $L^2$-estimate   for elliptic system, (\ref{lee2})
 and (\ref{qq1}) that\be\la{ua2-1} \ba
\norm[H^2]{F}+\norm[H^2]{\o} & \le
C\left(\norm[L^2]{F}+\norm[L^2]{\o}
 + \norm[L^2]{\rho\dot{u}}+ \norm[L^2]{\nabla(\rho\dot{u})} \right)\\
& \le C (1 + \norm[L^3]{\nabla\rho} \norm[L^6]{ \dot{u}} +
\norm[L^2]{\nabla\dot{u}}) \no & \le C (1+
\norm[L^2]{\nabla\dot{u}}), \ea \ee which, together with
(\ref{ua2}), Lemma \ref{le11}, and Gronwall's inequality, gives
directly \bnn \sup_{t\in[0,T]} {\left(\norm[L^2]{\nabla^2P }
+\norm[L^2]{\nabla^2 \rho } \right)}\le C. \enn Thus the proof of
Lemma \ref{le9-1} is completed.

\begin{lemma}\la{pe1}
The following estimates hold: \be\la{va5}
   \sup\limits_{0\le t\le T}\left(
   \|\n_t\|_{H^1}+\|P_t\|_{H^1}\right)
    + \int_0^T\left(\|\n_{tt}\|_{L^2}^2+\|P_{tt}\|_{L^2}^2\right)dt
\le C,
  \ee
\be\la{nq1}
   \sup\limits_{0\le t\le T}\int_{ }|\nabla u_t|^2dx
    + \int_0^T\int\rho u_{tt}^2dxdt
\le C.
  \ee

\end{lemma}
{\it Proof.} We first prove (\ref{va5}). One deduces from (\ref{p0})
and (\ref{qq1}) that \be \la{sp1} \|P_t\|_{L^2}\le
C\|u\|_{L^\infty}\|\nabla P\|_{L^2}+C\|\nabla u\|_{L^2}\le C.\ee
Differentiating (\ref{p0}) yields \bnn \nabla P_t+u\cdot\nabla\nabla
P+\nabla u\cdot\nabla P+\ga \nabla P {\rm div}u+\ga P  \nabla{\rm
div}u=0.\enn Hence, by (\ref{qq1}) and (\ref{va1}), one gets
\bn\la{sp2} \|\nabla P_t\|_{L^2}\le C\|u\|_{L^\infty}\|\nabla^2
P\|_{L^2}+C\|\nabla u\|_{L^3}\|\nabla P\|_{L^6}+C\|\nabla^2
u\|_{L^2}\le C.\en The combination of (\ref{sp1}) with (\ref{sp2})
implies \bn \la{sp3}\sup_{0\le t\le T}\|P_t\|_{H^1}\le C.\en Note
that $P_{tt}$ satisfies \be\la{s4} P_{tt} + \gamma P_t{\rm div}u +
\gamma P{\rm div}u_t + u_t\cdot\nabla P + u\cdot\nabla P_t = 0. \ee
Thus, one gets from (\ref{s4}) (\ref{sp3}) (\ref{qq1}) and
(\ref{ja1}) that \be \ba &\ia \|P_{tt}\|_{L^2}^2dt\no & \le C\ia
\left(\|P_t\|_{L^6}\|\nabla u\|_{L^3}+ \|\nabla
u_t\|_{L^2}+\|u_t\|_{L^6}\|\nabla P\|_{L^3}+\|\nabla
P_t\|_{L^2}\right)^2dt\\&\le C.\ea\ee One can handle $\n_t$ and
$\n_{tt}$ similarly. Thus (\ref{va5}) is proved.

Next, we prove (\ref{nq1}). Differentiating  $(\ref{a1})_2$  with
respect to $t,$ then multiplying the resulting equation  by
$u_{tt},$  one gets after integration by parts  that \be\la{sp9} \ba
& \frac{1}{2}\frac{d}{dt}\int \left(\mu|\nabla u_t|^2 + (\lambda +
\mu)({\rm div}u_t)^2\right)dx+\int_{ }\rho u_{tt}^2dx
\\
&=\frac{d}{dt}\left(-\frac{1}{2}\int_{ }\rho_t |u_t|^2 dx- \int_{
}\rho_t u\cdot\nabla u\cdot u_tdx+ \int_{ }P_t {\rm
div}u_tdx\right)\\&\quad + \frac{1}{2}\int_{ }\rho_{tt} |u_t|^2 dx+
\int_{ }(\rho_{t} u\cdot\nabla u )_t\cdot u_tdx-\int_{ }\rho
u_t\cdot\nabla u\cdot u_{tt}dx\\ &\quad- \int_{ }\rho u\cdot\nabla
u_t\cdot u_{tt}dx - \int_{ }P_{tt}{\rm div}u_tdx \\ &\triangleq
\frac{d}{dt}I_0+ \sum\limits_{i=1}^5I_i. \ea \ee

It follows from $(\ref{a1})_1,$  (\ref{qq1}), (\ref{va5}) and
(\ref{ja1}) that\be \ba \la{sp10}|I_0|& =\left|-\frac{1}{2}\int_{
}\rho_t |u_t|^2 dx- \int_{ }\rho_t u\cdot\nabla u\cdot u_tdx+
\int_{ }P_t {\rm div}u_tdx\right|\\ &\le \left|\int_{ } {\rm
div}(\n
u)|u_t|^2dx\right|+C\norm[L^3]{\rho_t}\norm[L^2]{u\cdot\nabla u}
\norm[L^6]{u_t}+C\|P_t\|_{L^2}\|\nabla u_t\|_{L^2}\\ &\le C \int_{
} \n |u||u_t||\nabla u_t| dx +C\|\nabla u_t\|_{L^2} \\ &\le C
\|u\|_{L^6}\|\n^{1/2} u_t\|_{L^2}^{1/2}\|u_t\|_{L^6}^{1/2}\|\nabla
u_t\|_{L^2} +C\|\nabla u_t\|_{L^2}\\ &\le \de\|\nabla
u_t\|_{L^2}^2+C_\de,\ea\ee
 \be \la{sp11}\ba
2|I_1|&=\left|\int_{ }\rho_{tt} |u_t|^2 dx\right|\\
& = \left|\int_{ }(\rho_tu + \rho u_t)\cdot\nabla( |u_t|^2)dx\right|\\
 & \le  C\left(\norm[L^3]{\rho_t}\norm[L^{\infty}]{u}
  +\norm[L^2]{\rho^{{1/2}}u_t}^{1/2}\|u_t\|_{L^6}^{1/2}\right)
  \|u_t\|_{L^6}
\norm[L^2]{\nabla u_t}\\
& \le C\norm[L^2]{\nabla u_t}^2 +C\norm[L^2]{\nabla u_t}^{5/2}\\
& \le C\norm[L^2]{\nabla u_t}^4 +C,
   \ea \ee
   and
\be \la{sp12}\ba
  |I_2|&=\left|\int_{ }\left(\rho_t u\cdot\nabla u \right)_t\cdot u_{t}dx
 \right|\\
& = \left|  \int_{ }\left(\rho_{tt} u\cdot\nabla u\cdot u_t +\rho_t
u_t\cdot\nabla u\cdot u_t+\rho_t u\cdot\nabla u_t\cdot
u_t\right)dx\right|\\ &\le   \norm[L^2]{\rho_{tt}}
\norm[L^3]{u\cdot\nabla
u}\norm[L^6]{u_t}+\norm[L^2]{\rho_t}\norm[L^3]{|u_t|^2}
\norm[L^6]{\nabla u} \\
&\quad+\norm[L^3]{\rho_t}\norm[L^{\infty}]{u}
\norm[L^2]{\nabla u_t}\norm[L^6]{u_t}\\
& \le C\norm[L^2]{\rho_{tt}}^2 + C\norm[L^2]{\nabla u_t}^2. \ea \ee
 Cauchy's inequality gives
\be\ba\la{sp13} |I_3|+|I_4|&= \left| \int_{ }\rho u_t\cdot\nabla
u\cdot u_{tt} dx\right| +\left| \int_{ }\rho u\cdot\nabla u_t\cdot
u_{tt} dx\right|\\& \le   C\|\n^{1/2}u_{tt}\|_{L^2}\left(
\|u_t\|_{L^6}\|\na u\|_{L^3}+\|u\|_{L^\infty}\|\na
u_t\|_{L^2}\right) \\& \le  \de \norm[L^2]{\rho^{{1/2}}u_{tt}}^2 +
C_\de\norm[L^2]{\nabla u_t}^2 , \ea\ee and \be\ba\la{sp15} |I_5|&=
\left|\int_{ }P_{tt}{\rm div}u_tdx\right|\\&\le
\norm[L^2]{P_{tt}}\norm[L^2]{{\rm div}u_t}\\& \le
C\norm[L^2]{P_{tt}}^2 + \norm[L^2]{\nabla u_t}^2. \ea\ee Due to the
regularity of the local solution, (\ref{ba1}), $t\nabla u_t\in
C([0,T_*];L^2)$. Thus \be \la{sp16}\ba \norm[L^2 ]{\nabla
u_t(\cdot,T_*/2)}  &\le \frac{2}{T_*}\norm[L^{\infty}(0,T_*;L^2
)]{t\nabla u_t}\\ &\le C, \ea\ee where $C$ may also depend  on
$\norm[L^2]{\nabla g}  .$

Collecting all the estimates (\ref{sp10})-(\ref{sp16}), one
deduces from (\ref{sp9}), (\ref{va5}), (\ref{ja1}) and Gronwall's
inequality that \be\la{sp17} \sup\limits_{T_*/2\le t\le T}\|\nabla
u_t\|_{L^2}+\int_{T_*/2}^T\int_{ }\n |u_{tt}|^2dxdt\le C.\ee
 On the other hand,
 (\ref{ba1}) gives \be\la{sp18} \sup\limits_{0\le t\le T_*/2 }\|\nabla
u_t\|_{L^2}+\int_0^{T_*/2} \int_{ }\n |u_{tt}|^2dxdt\le C.\ee The
combination of (\ref{sp17}) with (\ref{sp18}) yields (\ref{nq1})
immediately. This completes the proof of Lemma \ref{pe1}.

\begin{lemma}\la{pr3}
It holds that \be\la{y1}\ba \sup_{t\in[0,T]}\left(\norm[H^3]{
\rho-\tilde{\rho} } +\norm[H^3]{
  P-P(\tn )}
   \right)   \le C,\ea \ee \be\la{y2}\ba
\sup_{t\in[0,T]}\left(\|\na u_t\|_{L^2} + \norm[H^2]{\nabla u}
\right) +\ia \left(\|\na u\|_{H^3}^2+\|\na u_t\|_{H^1}^2\right)dt\le
C.\ea \ee
\end{lemma}

 {\it Proof.} It follows from  (\ref{nq1})  and
(\ref{qq1}) that \bnn \ba \|\nabla (\n \dot u) \|_{L^2}&\le
 \||\nabla \n | |  u_t|  \|_{L^2}+ \|\n \nabla   u_t  \|_{L^2}
 + \||\nabla \n|| u||\nabla u| \|_{L^2}\\ &
 + \|\n|\nabla  u|^2\|_{L^2}
 + \|  \n |u || \nabla^2 u| \|_{L^2}\\&\le
 \|\nabla \n \|_{L^3} \|  u_t  \|_{L^6}+ C\| \nabla   u_t  \|_{L^2}
 + C\| \nabla \n\|_{L^3}\| u\|_{L^\infty}\|\nabla u \|_{L^6}\\
 &\quad + C\| \nabla  u\|_{L^3}\| \nabla  u\|_{L^6}
 + C\|     u \|_{L^\infty}\| \nabla^2 u  \|_{L^2}\\ &\le C,\ea\enn
which together with (\ref{lee2}) gives \be\la{sp19} \sup_{0\le t\le
T} \|\n \dot u\|_{H^1}\le C.\ee

The standard $H^1$-estimate for elliptic system  (\ref{zp202}) gives
\be\la{sp20} \ba\|\nabla^2 u\|_{H^1} &\le C\|\mu\Delta
u+(\mu+\lambda)\nabla{\rm div}u \|_{H^1}\\&=C\|\n \dot u+\nabla
P\|_{H^1} \\&\le
 C (\|\n \dot u\|_{H^1}+ \|\nabla P\|_{H^1})\\&\le C,\ea\ee due to
$(\ref{a1})_2,$ (\ref{sp19}) and (\ref{va1}). As a consequence of
(\ref{qq1}) and (\ref{sp20}), one has \be\la{sp21} \sup\limits_{0\le
t\le T}\|\nabla  u\|_{H^2} \le
 C.\ee

Therefore, the standard $L^2$-estimate for elliptic system,
(\ref{qq1}), and Lemma \ref{pe1} yield  that
  \be\la{sp23}\ba\|\na^2u_t\|_{L^2}
 &\le C\|\mu\Delta u_t+(\mu+\lambda)\nabla{\rm div}u_t
 \|_{L^2}\\ &=\|\n  u_{tt}+\n_t u_t+\n_t u\cdot\nabla u
  +\n u_t\cdot\nabla u+\n u\cdot\nabla u_t+
  \nabla P_t\|_{L^2}   \\
&\le C\left(\|\n  u_{tt}\|_{L^2}+ \|\n_t\|_{L^3}
\|u_t\|_{L^6}+\|\n_t\|_{L^3}\| u\|_{L^\infty}\|\nabla
u\|_{L^6}\right)\\&\quad
  +C\left(\| u_t\|_{L^6}\|\nabla u\|_{L^3}+ \| u\|_{L^\infty}
  \|\nabla u_t\|_{L^2}+\|
  \nabla P_t\|_{L^2}\right)\\ &\le C\|\n  u_{tt}\|_{L^2} +C,\ea \ee
which, together with (\ref{nq1}), implies \be\la{sp24} \ia
\|\nabla u_t\|_{H^1}^2dt\le C . \ee

Applying the standard $H^2$-estimate for elliptic system
(\ref{zp202}) again leads to \be \la{sp38}\ba \|\na^2 u\|_{H^2}&\le
C\|\mu\Delta u+(\mu+\lambda)\nabla{\rm div}u\|_{H^2}\\&\le C\| \n
\dot u \|_{H^2}+C\|\na  P
 \|_{H^2}\\ & \le C +C \|\na  u_t\|_{H^1}+C\|\na^3 P
 \|_{L^2},\ea\ee
where one has used (\ref{sp19}) and the following simple facts: \be
\la{sp25}\ba \|\nabla^2(\n u_t)\|_{L^2}&\le C\left(
\||\na^2\n||u_t|\|_{L^2}+\||\na\n||\na u_t|\|_{L^2}+\| \na^2 u_t
\|_{L^2}\right)\no & \le C\left( \| \na^2\n\|_{L^2}\|\na
u_t\|_{H^1}+\|\na\n\|_{L^3}\|\na u_t \|_{L^6}+\| \na^2 u_t
\|_{L^2}\right)\\ & \le C +C \|\na u_t\|_{H^1},\ea\ee and
\be\la{sq26}\ba \|\na^2(\n u\cdot\na u)\|_{L^2}&\le C\left(
\||\na^2(\n u)||\na u|\|_{L^2} + \||\na (\n u)||\na^2 u|\|_{L^2} +
\| \na^3 u\|_{L^2} \right)\no &\le C\left( 1+\| \na^2(\n
u)\|_{L^2}\|\na u\|_{H^2} + \|\na (\n u)\|_{L^3}\|\na^2 u\|_{L^6}
\right)\\ &\le C\left( 1+\| \na^2 \n  \|_{L^2} \|u\|_{L^\infty}
+\| \na \n \|_{L^6} \| \na u \|_{L^3}+\| \na^2 u \|_{L^2}\right)\\
&\le C,\ea\ee due to (\ref{va1}) and (\ref{sp21}). By using
(\ref{sp21}), (\ref{sp38}), and (\ref{va1}), one may get that
 \be\ba &\left(\|\na^3 P\|_{L^2}^2\right)_t\no&\le
C\left(\||\na^3u||\na P|\|_{L^2}+ \||\na^2u||\na^2 P|\|_{L^2}+
\||\na u||\na^3 P|\|_{L^2}+ \| \na^4u
\|_{L^2}\right)\|\na^3P\|_{L^2} \\&\le C\left(\| \na^3u\|_{L^2}\|\na
P \|_{H^2}+ \| \na^2u\|_{L^3}\|\na^2 P \|_{L^6}+ \| \na
u\|_{L^\infty}\|\na^3 P \|_{L^2}\right)\|\na^3P\|_{L^2}\\
&\quad+C\left(
 1+\|\na^2u_t \|_{L^2}+ \| \na^3P
\|_{L^2}\right) \| \na^3P \|_{L^2} \\ &\le C+C \| \na u_t
\|^2_{H^1}+C \| \na^3P \|^2_{L^2},  \ea\ee which, together with
Gronwall's inequality and (\ref{sp24}), yields that \be\la{sp26}
\sup\limits_{0\le t\le T}\|\nabla^3  P\|_{L^2} \le C.\ee Collecting
all these estimates (\ref{sp24})-(\ref{sp26}) and (\ref{va1}) shows
 \be\la{sp27}
\sup\limits_{0\le t\le T}\| P-P(\tn)\|_{H^3} +\int_0^T\|\na
u\|^2_{H^3}dt\le
 C.\ee
It is easy to check similar arguments work for $\n-\tn $ by using
(\ref{sp27}). Hence, \be\la{sp28} \sup\limits_{0\le t\le T}\|
\n-\tn\|_{H^3} \le
 C.\ee
 Combing (\ref{sp27}) with (\ref{sp28}) shows (\ref{y1}).
Estimate (\ref{y2}) thus follows from (\ref{nq1}), (\ref{sp21}),
(\ref{sp24}), and (\ref{sp27}). Hence the proof of Lemma \ref{pr3}
is finished.

\begin{lemma}\la{sq90} For any $\tau\in (0,T),$ there exists some positive constant
$C(\tau)$ such that
 \be \la{y3}\sup_{\tau\le t\le T}\left(\|\na u_t\|_{H^1}
 +\|\na^4 u \|_{L^2}\right)
 +\int_{\tau}^T\int_{ }|\nabla u_{tt}|^2dxdt\le
C(\tau).\ee

\end{lemma}

{\it Proof.} Differentiate $(\ref{a1})_2$ with respect to $t$ twice
to get \be\la{sp30}\ba &\n u_{ttt}+\n u\cdot\na u_{tt}-\mu\Delta
u_{tt}-(\mu+\lambda)\nabla{\rm div}u_{tt}\\&= 2{\rm div}(\n u)u_{tt}
+{\rm div}(\n u)_{t}u_t-2(\n u)_t\cdot\na u_t-(\n_{tt} u+2\n_t u_t)
\cdot\na u\\& \quad- \n u_{tt}\cdot\na u-\na P_{tt}.
 \ea\ee
Multiplying (\ref{sp30}) by $u_{tt}$ and then integrating the
resulting equation over ${\r^3} ,$ one gets after integration by
parts that \be \la{sp31}\ba &\frac{1}{2}\frac{d}{dt}\int_{ }\n
|u_{tt}|^2dx+\int_{ }\left(\mu|\na u_{tt}|^2+(\mu+\lambda)({\rm
div}u_{tt})^2\right)dx \\&=-4\int_{ }  u^i_{tt}\n u\cdot\na
 u^i_{tt} dx-\int_{ }(\n u)_t\cdot \left[\na (u_t\cdot u_{tt})+2\na
u_t\cdot u_{tt}\right]dx\\&\quad -\int_{
}(\n_{tt}u+2\n_tu_t)\cdot\na u\cdot u_{tt}dx-\int_{ }   \n
u_{tt}\cdot\na u\cdot  u_{tt} dx+\int_{ } P_{tt}{\rm
div}u_{tt}dx\\&\triangleq\sum_{i=1}^5J_5.\ea\ee

We estimate each $J_i(i=1,\cdots,5)$ as follows:

 H\"{o}lder's
inequality gives\be \la{sp32} \ba |J_1|&\le
C\|\n^{1/2}u_{tt}\|_{L^2}\|\na u_{tt}\|_{L^2}\| u \|_{L^\infty}\\
&\le \de \|\na u_{tt}\|_{L^2}^2+C_\de\|\n^{1/2}u_{tt}\|^2_{L^2}
.\ea\ee It follows from (\ref{ja1}), (\ref{va5}), (\ref{nq1}), and
(\ref{qq1}) that \be \la{sp33}\ba |J_2|&\le C\left(\|\n
u_t\|_{L^3}+\|\n_t u\|_{L^3}\right)\left(\| u_{tt}\|_{L^6}\| \na
u_t\|_{L^2}+\| \na u_{tt}\|_{L^2}\| u_t\|_{L^6}\right)\\&\le
C\left(\|\n^{1/2} u_t\|^{1/2}_{L^2}\|u_t\|^{1/2}_{L^6}+\|\n_t
\|_{L^6}\| u\|_{L^6}\right)  \| \na u_{tt}\|_{L^2} \\ &\le \de
\|\na u_{tt}\|_{L^2}^2+C_\de,\ea\ee

\be  \la{sp34}\ba |J_3|&\le C\left(\|\n_{tt}\|_{L^2}
\|u\|_{L^\infty}\|\na u\|_{L^3}+\|\n_{
t}\|_{L^6}\|u_{t}\|_{L^6}\|\na u \|_{L^2}\right)\|u_{tt}\|_{L^6} \\
&\le \de \|\na u_{tt}\|_{L^2}^2+C_\de\|\n_{tt}\|_{L^2}^2,\ea\ee and
\be  \la{sp36}\ba |J_4|+|J_5|&\le C\|\n u_{tt}\|_{L^2} \|\na
u\|_{L^3}\|u_{tt}\|_{L^6} +C \|P_{tt}\|_{L^2}\|\na
u_{tt}\|_{L^2}\\
&\le \de \|\na u_{tt}\|_{L^2}^2+C_\de\|\n^{1/2}u_{tt}\|^2_{L^2}
+C_\de\|P_{tt}\|^2_{L^2}. \ea\ee For any $\tau\in (0,T_*),$ since
$t^{1/2}\sqrt{\n}u_{tt}\in L^{\infty}(0,T_*;L^2)$ by (\ref{ba1}),
there exists some $t_0\in (\tau/2,\tau)$ such that \be\la{sp37}\ba
\int_{ }\n |u_{tt}|^2dx(t_0)&\le \frac{1}{ t_0
}\|t^{1/2}\sqrt{\n}u_{tt}\|^2_{L^\infty(0,T_*;L^2)}\\&\le C(\tau
).\ea\ee Substituting (\ref{sp32})-(\ref{sp36}) into (\ref{sp31})
and choosing $\de$ suitably small, one obtains by using (\ref{va5})
(\ref{sp37}) and Gronwall's inequality that \bnn \sup_{t_0\le t\le
T}\int_{ }\n |u_{tt}|^2dx+\int_{t_0}^T\int_{ }|\nabla
u_{tt}|^2dxdt\le C(\tau),\enn which, together with (\ref{sp23}) and
(\ref{nq1}), yields that
 \be \la{sp40}\sup_{\tau\le t\le T}\|\na u_t\|_{H^1}
 +\int_{\tau}^T\int_{ }|\nabla u_{tt}|^2dxdt\le
C(\tau),\ee due to $t_0<\tau.$ Now, (\ref{y3}) follows from
(\ref{sp38}), (\ref{sp40}), and (\ref{y1}). We finish the proof of
Lemma \ref{sq90}.

\section{\la{se4}Proof of  Theorems  \ref{th1} and \ref{th2}}

With all the a priori estimates in Section \ref{se3} at hand, we are
ready to prove the main results of this paper in this section.

{\it Proof of Theorem \ref{th1}.} By Lemma \ref{th0}, there exists a
$T_*>0$ such that the Cauchy problem (\ref{a1}), (\ref{h1}),
(\ref{h2}) has a unique classical solution $(\rho,u)$ on $\r^3\times
(0,T_*]$. We will use the a priori estimates, Proposition \ref{pr1}
and Lemmas \ref{pr3} and \ref{sq90}, to extend the local classical
solution $(\rho,u)$ to all time.

First,  it follows from (\ref{As1}), (\ref{As2}), (\ref{uv4}) and
(\ref{h7})    that
$$ A_1(0)+A_2(0)=0, \quad A_3(0) \leq C_0^{\de_0}, \quad \rho_0\leq
\bar{\rho},$$ due to  $C_0\le \ve.$ Therefore, there exists a
$T_1\in(0,T_*]$ such that (\ref{z1}) holds for $T=T_1$.

Next, we set \bn \la{s1}T^*=\sup\{T\,|\,{\rm (\ref{z1}) \
holds}\}.\en Then $T^*\geq T_1>0$. Hence, for any $0<\tau<T\leq T^*$
with $T$ finite, it follows from Lemmas \ref{pr3} and \ref{sq90}
that
 \be \la{sp43}\na u_t, \na^3u\in C([\tau ,T];L^2\cap L^4),\quad
 \na u,\na^2u \in C\left([\tau ,T];L^2\cap
 C\left(\overline{{\r^3}}\right)\right),\ee where we have used the standard
embedding
$$L^\infty(\tau ,T;H^1)\cap H^1(\tau ,T;H^{-1})\hookrightarrow
C\left([\tau ,T];L^q\right),\quad\mbox{ for any } q\in [2,6).  $$
Due to (\ref{ja1}), (\ref{nq1}), and (\ref{y3}), one can get \be\ba
&\int_{\tau}^T \|(\n |u_t|^2)_t\|_{L^1}dt\no &\le
\int_{\tau}^T\left(\|  \n_t  |u_t|^2 \|_{L^1}+2\|  \n  u_t\cdot
u_{tt} \|_{L^1}\right)dt\\ &\le C\int_{\tau}^T \left( \| \n|\div u|
|u_t|^2 \|_{L^1}+\|  |u||\na \n| |u_t|^2 \|_{L^1}+ \| \n^{1/2} u_t
\|_{L^2}\|\n^{1/2}u_{tt} \|_{L^2}\right)dt\\ &\le C\int_{\tau}^T
\left( \| \n |u_t|^2 \|_{L^1}\|\na u\|_{L^\infty}+\|  u\|_{L^6}\|\na
\n\|_{L^2} \|u_t  \|^2_{L^6}+  \|\n^{1/2}u_{tt} \|_{L^2}\right)dt\\
&\le C,\ea\ee which yields \bnn \n^{1/2}u_t\in
C([\tau,T];L^2).\enn This, together with (\ref{sp43}), gives \be
\la{n20} \n^{1/2}\dot u,\na \dot u\in C([\tau,T];L^2).\ee

Next, we claim that \be \la{s2}T^*=\infty.\ee Otherwise,
$T^*<\infty$. Then by Proposition \ref{pr1}, (\ref{z2}) holds for
$T=T^*$. It follows from Lemmas \ref{pr3} and \ref{sq90} and
(\ref{n20}) that $(\n(x,T^*),u(x,T^*))$ satisfies
(\ref{co1})-(\ref{co2})  except $ u(\cdot,T^*)\in \dot H^\beta,$
where  $g(x)\triangleq\dot u(x, T^*),\,\,x\in \r^3.$ Thus, Lemma
\ref{th0} implies that there exists some $T^{**}>T^*$, such that
(\ref{z1}) holds for $T=T^{**}$, which contradicts (\ref{s1}).
Hence, (\ref{s2}) holds. Lemmas \ref{th0}, \ref{pr3} and \ref{sq90}
and (\ref{sp43}) thus show that $(\rho,u)$ is in fact the unique
classical solution defined on $\r^3\times(0,T]$ for any
$0<T<T^*=\infty$.

Finally, to finish the proof of Theorem \ref{th1}, it remains to
prove (\ref{h11}).

Multiplying (\ref{a95}) by $4 (\pp)^3$ and integrating the resulting
equality over ${\r^3}$, one has \be\ba&
\left(\|\pp\|_{L^4}^4\right)'(t)\no&=-(4\ga-1)\int (\pp)^4\div
udx-\ga\int P(\tn)(\pp)^3\div udx,\ea\ee which yields that
\be\la{n3}\ba \int_1^\infty\left|\left(\|\pp\|_{L^4}^4\right)'(t)
\right|dt&\le C\int_1^\infty\left(\|\pp\|_{L^4}^4 + \|\na
u\|_{L^4}^4\right)dt\le C,\ea\ee due to (\ref{j29}). Combining
(\ref{j29}) with (\ref{n3}) leads to \bnn\lim_{t\rightarrow
\infty}\|\pp\|_{L^4}=0,\enn which together with (\ref{a16}) implies
\bnn  \lim_{t\rightarrow\infty}\int |\rho-\tn|^q dx = 0, \enn for
all $q$ satisfying (\ref{eq1}). Note that (\ref{a16}) and (\ref{g1})
imply \bnn \int\n^{1/2}|u|^4dx\le \left(\int\n
|u|^2dx\right)^{1/2}\|u\|_{L^6}^3\le C\|\na u\|_{L^2}^3. \enn Thus
(\ref{h11}) follows provided that
  \be\la{n8}\lim_{t\rightarrow \infty}\|\na
u\|_{L^2}=0.\ee   Setting \bnn I(t)\triangleq \frac{\mu
}{2}\|\nabla u\|_{L^2}^2+\frac{\lambda+\mu
}{2}\|\text{div}u\|_{L^2}^2, \enn choosing $m=0$ in (\ref{m0}),
and using (\ref{m2}) and (\ref{m3}), one has \be\la{n6} |I'(t)|
\le C\int\n |\dot u|^2dx+C\|\na u\|_{L^3}^3+CC_0^{1/2}\|\na \dot
u\|_{L^2},\ee where one has used the following simple estimate:
\be \ba |M_1| & = \left| \int \dot{u}\cdot\nabla Pdx\right|\no & =
\left| \int (\pp) \div\dot{u}
 dx\right|\\ &\le CC_0^{1/2}\|\na\dot u\|_{L^2}.\ea\ee
We thus deduce from (\ref{n6}), (\ref{h27}), and (\ref{j29}) that
\bnn\ba \int_1^\infty|I'(t)|^2dt&\le
C\int_1^\infty\left(\|\n^{1/2}\dot u\|_{L^2}^4+\|\na
u\|^2_{L^2}\|\na u\|_{L^4}^4+\|\na \dot u\|_{L^2}^2\right)dt\\&\le
C\int_1^\infty\left(\|\n^{1/2}\dot u\|_{L^2}^2+ \|\na
u\|_{L^4}^4+\|\na \dot u\|_{L^2}^2\right)dt
\\&\le C,\ea\enn which, together with \bnn\int_1^\infty |I(t)|^2dt\le C
\int_1^\infty \|\na u\|_{L^2}^2dt\le C,\enn implies (\ref{n8}).
The proof of Theorem \ref{th1} is finished.

{\it Proof of  Theorem   \ref{th2}. } The proof is similar to
 that of Theorem 1.2 in \cite{lx}. We just sketch it here.

Otherwise, there exist some constant $C_1>0 $ and a subsequence
$\left\{t_{n_j}\right\}_{j=1}^\infty ,$ $t_{n_j}\rightarrow \infty$
such that $\left\|\nabla \rho
\left(\cdot,t_{n_j}\right)\right\|_{L^{r} }\le C_1.$ Hence, the
Gagliardo-Nirenberg inequality (\ref{g2}) yields that there exists
some positive constant $C$ independent of $t_{n_j} $  such that for
$a=r/(2r-3)\in (0,1), $
 \bn\la{eq3}
\lefteqn{\left\|\rho (x,t_{n_j})-\tn
\right\|_{C\left(\ol{{\r^3}}\right)}}\no &&\le C \left\|\nabla
 \rho  (x,t_{n_j}) \right\|^a_{L^r}
\left\|\rho  (x,t_{n_j})-\tn \right\|^{1-a}_{L^3}
 \no&&\le C C_1^a\left\|\rho
 (x,t_{n_j})-\tn\right\|^{1-a}_{L^3}
.\en    Due to (\ref{h11}), the right hand side of (\ref{eq3})
goes to $0$ as $t_{n_j}\rightarrow \infty.$ Hence,
\bn\la{e9}\left\|\rho
 (x,t_{n_j})-\tn
\right\|_{C\left(\ol{{\r^3}}\right)}\rightarrow 0\mbox{ as
}t_{n_j}\rightarrow \infty.\en

On the other hand, since $(\n,u)$ is a classical solution  satisfying
(\ref{h9}),   there exists a unique particle path $x_0(t)$ with
$x_0(0)=x_0$ such that
$$\rho (x_0(t),t)\equiv 0 \mbox{  for all }  t\ge 0. $$ So, we conclude
from this identity that \bnn\left\|\rho
 (x,t_{n_j})-\tn\right\|_{C\left(\ol\O\right)} \ge
 \left|\rho
(x_0(t_{n_j}),t_{n_j})-\tn\right|  \equiv  \tn
>0,\enn which contradicts (\ref{e9}). This completes the proof of
Theorem 1.2.



\begin {thebibliography} {99}

\bibitem{bkm} Beal, J. T., Kato, T., Majda. A.:
Remarks on the breakdown of smooth solutions for the 3-D Euler
equations.  Commun. Math. Phys {\bf 94}, 61-66 (1984)

\bibitem{bl} Bergh, J.,    Lofstrom, J.: {\it Interpolation spaces, An
introduction,}   Berlin-Heidelberg-New York:Springer-Verlag, (1976)

\bibitem{K1} Cho, Y., Choe, H. J.,   Kim, H.:
Unique solvability of the initial boundary value problems for
compressible viscous fluid. J. Math. Pures Appl. {\bf 83},
 243-275 (2004)

\bibitem{cj} Cho, Y.,  Jin, B.J.:  Blow-up of viscous heat-conducting
compressible flows, J. Math. Anal. Appl. {\bf 320}(2),
 819-826 (2006)


\bibitem{K3} Cho, Y.,   Kim, H.:
On classical solutions of the compressible Navier-Stokes equations
with nonnegative initial densities. Manuscript Math. {\bf 120},
91-129 (2006)


\bibitem{K2} Choe, H. J.,    Kim, H.:
Strong solutions of the Navier-Stokes equations for isentropic
compressible fluids. J. Differ. Eqs. {\bf 190}, 504-523 (2003)


\bibitem{F1} Feireisl, E., Novotny, A., Petzeltov\'{a}, H.: On the existence of globally defined weak solutions to the
Navier-Stokes equations. J. Math. Fluid Mech. {\bf 3}(4), 358-392
(2001)

\bibitem{fk} Fujita, H., Kato, T.: On the Navier-Stokes initial value problem I.
Archiv Rat. Mech. Anal. {\bf 16}, 269-315 (1964)

\bibitem{Hof} Hoff, D.:
Global existence for 1D, compressible, isentropic Navier-Stokes
equations with large initial data. Trans. Amer. Math. Soc. {\bf
303}(1), 169-181 (1987)

\bibitem{H3}Hoff, D: Global solutions of the Navier-Stokes equations
 for multidimensional compressible flow with discontinuous initial data.
J. Differ. Eqs.  {\bf 120}(1), 215-254 (1995)

\bibitem{Hof2}Hoff, D.:
Strong convergence to global solutions for multidimensional flows of
compressible, viscous fluids with polytropic equations of state and
discontinuous initial data.  Arch. Rational Mech. Anal.  {\bf 132},
1-14 (1995)

\bibitem{Ho3}Hoff, D.: Compressible flow in a half-space with Navier boundary
  conditions. J. Math. Fluid Mech. {\bf 7}(3), 315-338 (2005)


\bibitem{hof2002}Hoff, D.: Dynamics of singularity surfaces for compressible,
viscous flows in two space dimensions. Comm. Pure Appl. Math. {\bf
55}(11), 1365-1407 (2002)

 \bibitem{hs}  Hoff, D.,   Santos, M. M.:
Lagrangean structure and propagation of singularities in
multidimensional compressible flow.  Arch. Rational Mech. Anal.
{\bf 188}(3), 509-543 (2008)

 \bibitem{ht}
Hoff, D., Tsyganov, E.: Time analyticity and backward uniqueness of
weak solutions of the Navier-Stokes equations of multidimensional
compressible flow. J. Differ. Eqs.  {\bf 245}(10) 3068-3094  (2008)






\bibitem{hlx1} Huang, X. D., Li, J., Xin Z. P.:
Blowup criterion for viscous barotropic flows with vacuum states.
Commun. Math. Phys., In press.

\bibitem{hlx} Huang, X. D., Li, J., Xin Z. P.:
Serrin type criterion for the three-dimensional compressible flows.
Preprint


\bibitem{hlx3} Huang, X. D., Li, J., Luo, Z.,  Xin Z. P.:
Global existence and blowup phenomena for
  smooth solutions to  the two-dimensional compressible flows.
Preprint

\bibitem{hlx4} Huang, X. D., Li, J., Xin Z. P.:
Global well-posedness  for
  classical solutions to  the  multi-dimensional isentropic
   compressible
 Navier-Stokes system with vacuum on bounded domains. In preparation, 2010.

\bibitem{hx2} Huang, X. D.,  Xin, Z. P.:
A blow-up criterion for classical solutions to the compressible
Navier-Stokes equations,  Sci. in China,     {\bf 53}(3),  671-686
(2010)


\bibitem{Kaz} Kazhikhov, A. V.,  Shelukhin, V. V.:
Unique global solution with respect to time of initial-boundary
value problems for one-dimensional equations of a viscous gas.
Prikl. Mat. Meh.  {\bf 41}, 282-291 (1977)


\bibitem{kato} Kato, T.: Strong $L^{p}$-solutions of the Navier-Stokes
  equation in $R^{m}$, with applications
  to weak solutions. Math. Z. {\bf 187}(4),  471-480  (1984)

\bibitem{koch}Koch, H., Tataru, D.: Well-posedness for the Navier-Stokes
equations. Adv. Math. {\bf 157}(1), 22-35 (2001)

\bibitem{la}
Ladyzenskaja, O. A., Solonnikov,  V. A.,   Ural'ceva, N. N.:\emph{
Linear and quasilinear equations of parabolic type,} American
Mathematical Society, Providence, RI (1968)

\bibitem{lx}Li, J., Xin, Z.: Some uniform estimates and blowup behavior of
global strong solutions to the Stokes approximation equations for
two-dimensional compressible flows. J. Differ. Eqs.  {\bf 221}(2),
  275-308 (2006).

\bibitem{L1} Lions, P. L.: \emph{Mathematical topics in fluid mechanics}. Vol. {\bf 2}. Compressible models. New York: Oxford
University Press  (1998)


\bibitem{M1} Matsumura, A.,   Nishida, T.:  The initial value problem for the equations of motion of viscous and heat-conductive
gases. J. Math. Kyoto Univ. {\bf 20}(1), 67-104 (1980)



\bibitem{Na} Nash, J.: Le probl\`{e}me de Cauchy pour les \'{e}quations
diff\'{e}rentielles d'un fluide g\'{e}n\'{e}ral. Bull. Soc. Math.
France. {\bf 90},487-497 (1962)

\bibitem{R}Rozanova, O.:  Blow up of smooth solutions to the compressible
Navier-Stokes equations with the data highly decreasing at infinity,
J. Differ. Eqs.  {\bf 245},  1762-1774 (2008)

\bibitem{S2} Salvi,R.,  Straskraba, I.:
Global existence for viscous compressible fluids and their behavior
as $t\rightarrow \infty$. J. Fac. Sci. Univ. Tokyo Sect. IA. Math.
{\bf 40}, 17-51 (1993)

\bibitem{Ser1} Serre, D.:
Solutions faibles globales des \'equations de Navier-Stokes pour un
fluide compressible. C. R. Acad. Sci. Paris S\'er. I Math.
 {\bf 303}, 639-642 (1986)

\bibitem{Ser2} Serre, D.:
Sur l'\'equation monodimensionnelle d'un fluide visqueux,
compressible et conducteur de chaleur. C. R. Acad. Sci. Paris S\'er.
I Math. {\bf 303}, 703-706 (1986)

\bibitem{se1} Serrin, J.: On the uniqueness of compressible fluid motion,
Arch. Rational. Mech. Anal. {\bf 3}, 271-288 (1959)



\bibitem{X1} Xin, Z. P.:
Blowup of smooth solutions to the compressible {N}avier-{S}tokes
equation with compact density. Comm. Pure Appl. Math.   {\bf 51},
229-240 (1998)



\bibitem{zl1}Zlotnik, A. A.:  Uniform estimates and stabilization of symmetric
solutions of a system of quasilinear equations.   Diff. Equations,
 {\bf 36},  701-716(2000)
\end {thebibliography}
\end{document}